\documentclass[11pt,a4paper,abstract=on]{scrartcl}

\usepackage[utf8]{inputenc}
\usepackage[english]{babel}

\usepackage[paper=a4paper,left=25mm,right=25mm,top=25mm,bottom=30mm]{geometry}
\usepackage{amsmath,amssymb,amsthm,mathrsfs,bbm,amsfonts,bm}
\usepackage{color,graphicx,cases,caption,enumerate}
\usepackage{float,multicol,authblk}
\usepackage[dvipsnames]{xcolor}
\usepackage{colortbl}
\usepackage{booktabs}% http://ctan.org/pkg/booktabs
\usepackage{multirow}% http://ctan.org/pkg/multirow
\usepackage{subcaption}
\usepackage{algorithmic}
\usepackage[linesnumbered,ruled]{algorithm2e}
\usepackage{setspace}
\usepackage{booktabs}
\usepackage{bbm}
\usepackage{capt-of}
\usepackage{array}
\usepackage[title]{appendix}
\newcolumntype{C}[1]{>{\centering\let\newline\\\arraybackslash\hspace{0pt}}m{#1}}
\usepackage{natbib}
\usepackage{placeins} % --> \FloatBarrier
\usepackage[breaklinks, colorlinks=true, citecolor=black, urlcolor=black, linkcolor=black]{hyperref}

\newcommand{\tabref}{Table \ref}
\newcommand{\figref}{Figure \ref}
\newcommand{\secref}{Section \ref}
\newcommand{\one}{\mathbbm{1}}

\newcommand{\nout}{n_\text{out}}

\title{Generative machine learning methods for multivariate ensemble post-processing}
\author[1]{Jieyu Chen}
\author[2]{Tim Janke}
\author[2]{Florian Steinke}
\author[1,3]{Sebastian Lerch}
\affil[1]{Karlsruhe Institute of Technology}
\affil[2]{Technical University of Darmstadt}
\affil[3]{Heidelberg Institute for Theoretical Studies}

\date{\today}

\begin{document}

\maketitle

\begin{abstract}
\noindent
 Ensemble weather forecasts based on multiple runs of numerical weather prediction models typically show systematic errors and require post-processing to obtain reliable forecasts. 
 Accurately modeling multivariate dependencies is crucial in many practical applications, and various approaches to multivariate post-processing have been proposed where ensemble predictions are first post-processed separately in each margin and multivariate dependencies are then restored via copulas. These two-step methods share common key limitations, in particular the difficulty to include additional predictors in modeling the dependencies.
 We propose a novel multivariate post-processing method based on generative machine learning to address these challenges. In this new class of nonparametric data-driven distributional regression models, samples from the multivariate forecast distribution are directly obtained as output of a generative neural network. The generative model is trained by optimizing a proper scoring rule which measures the discrepancy between the generated and observed data, conditional on exogenous input variables. Our method does not require parametric assumptions on univariate distributions or multivariate dependencies and allows for incorporating arbitrary predictors.
 In two case studies on multivariate temperature and wind speed forecasting at weather stations over Germany, our generative model shows significant improvements over state-of-the-art methods and particularly improves the representation of spatial dependencies.
\end{abstract}

\section{Introduction}\label{intro}

Most weather forecasts today are based on ensemble simulations of numerical weather prediction (NWP) models. 
Despite substantial improvements over the past decades \citep{BauerEtAl2015}, these ensemble predictions continue to exhibit systematic errors such as biases, and typically fail to correctly quantify forecast uncertainty.
These systematic errors can be corrected by statistical post-processing, the application of which has become standard practice in research and operations.
Over the past years, a focal point of research interest has been the use of modern machine learning (ML) methods for post-processing, where random forest or neural network models enable the incorporation of arbitrary input predictors and have demonstrated superior forecast performance \citep{TaillardatEtAl2016,rasp2018neural}. For a general overview of recent developments, we refer to \citet{vannitsem2021statistical} and \citet{Haupt2021}.

While much of the research interest in post-processing has been focused on univariate methods, many practical applications require accurate models of spatial, temporal, or inter-variable dependencies \citep{schefzik2013uncertainty}.
Key examples include energy forecasting \citep{PinsonMessner2018,WorsnopEtAl2018}, air traffic management \citep{ChaloulosLygeros2007} and hydrological applications \citep{ScheuererEtAl2017}.
While the spatial, temporal, or inter-variable dependencies are present in the raw ensemble predictions from the NWP model, they are lost if univariate post-processing methods are applied separately in each margin (i.e., at each location, time step and for each target variable), which corresponds to implicitly assuming independence across space, time and variables. 

Over the past decade, various multivariate post-processing methods have been proposed, see \citet{schefzik2018ensemble} and \citet{vannitsem2021statistical} for general overviews. 
The vast majority of multivariate post-processing methods follows a two-step strategy\footnote{
	There exist examples of directly fitting specific multivariate probability distributions, which are mostly used in low-dimensional settings or if a specific structure can be chosen for the application at hand, with examples focusing on spatial \citep{FeldmannEtAl2015}, temporal \citep{MuschinskiEtAl2022} and inter-variable \citep{SchuhenEtAl2012,BaranMoeller2015,LangEtAl2019} dependencies being available. In addition, there are alternative approaches such as member-by-member post-processing \citep{vanSchaeybroeckVannitsem2015} which inherently preserve dependencies present in the raw ensemble predictions.
}.
In the first step, univariate post-processing methods are applied independently in all dimensions to obtain calibrated marginal probability distributions. Samples are then generated from these marginal predictive distributions obtained after post-processing. 
In the second step, the univariate sample values are re-arranged according to a specific multivariate dependence template with the aim of restoring the multivariate dependencies that are lost in the first step.
From a mathematical perspective, this can be interpreted as the application of a (parametric or non-parametric) copula, i.e., a multivariate cumulative distribution function (CDF) with standard uniform marginal distributions \citep{Nelsen2006}. 

In most popular copula-based approaches to multivariate post-processing, the copula $C$ is chosen to be either the parametric Gaussian copula (in the Gaussian copula approach \citep[GCA;][]{moller2013multivariate,PinsonGirard2012}), or a non-parametric empirical copula induced by a pre-specified dependence template based on the raw ensemble predictions (in the ensemble copula coupling \citep[ECC;][]{schefzik2013uncertainty} approach) or past observations \citep[in the Schaake shuffle approach proposed by][]{clark2004schaake}.
More advanced variants of these methods which aim to better incorporate structure in forecast error autocorrelations \citep{BouallegueEtAl2016} or optimize the selection of the dependence template based on similarity \citep{Schefzik2017,ScheuererEtAl2017} have been proposed over the past years.
Several comparative studies of multivariate post-processing methods based on simulated and real-world data are available \citep{wilks2015multivariate,lerch2020simulation,PerroneEtAl2020,LakatosEtAl2022}.
Overall, findings from these studies indicate that there is no consistently best approach across all settings, and that 
the observed differences in predictive performance depend on the misspecifications of the raw ensemble predictions, but tend to be minor and not statistically significant. 

All these state-of-the-art two-step methods for multivariate post-processing share common key limitations. 
Perhaps most importantly, there is no straightforward way to include additional predictors beyond ensemble forecasts or past observations of the target variable in the second step of imposing the dependence template onto the samples from the univariate post-processed forecast distributions.
Incorporating additional predictors has proven to be a key aspect in the substantial improvements in predictive performance observed for ML-based univariate post-processing models \citep{TaillardatEtAl2016,rasp2018neural,vannitsem2021statistical}, and it seems reasonable to expect similar beneficial effects for modeling multivariate dependencies.
Further, the number of samples that can be obtained from the multivariate forecast distribution in the two-step post-processing methods is limited by the number of ensemble predictions (in the ECC approach), or the number of suitable past observations (in case of the Schaake shuffle) which might be disadvantageous for reliably representing and predicting multivariate extreme events.

To overcome these challenges, we propose a novel nonparametric multivariate post-processing method based on generative ML approaches. 
In this new class of data-driven multivariate distributional regression models, samples from the multivariate forecast distribution are directly obtained as output of a generative deep neural network which allows for incorporating arbitrary input predictors including NWP-based ensemble predictions and additional exogenous variables.
Our generative model circumvents the two-step structure of the state-of-the-art multivariate post-processing approaches and aims to simultaneously correct systematic errors in the marginal distributions and the multivariate dependence structure without requiring parametric assumptions. 

There has recently been an increase in research activity on generative machine learning methods for weather modeling, in particular in the context of downscaling precipitation forecasts \citep{leinonen2020stochastic,HarrisEtAl2022,HessEtAl2022,PriceRasp2022} and nowcasting \citep{ravuri2021skilful}, where several approaches based on generative adversarial networks (GANs) have been proposed. 
While GANs are particularly useful for generating realistic images and have shown some success in a post-processing application to total cloud cover forecasts \citep{DaiHemri2021}, their training is often challenging \citep{Gui2022}. 
We consider a conceptually simpler class of scoring rule-based generative models \citep{li2015generative, dziugaite2015training}, extending recent work in probabilistic model averaging in energy forecasting \citep{janke2020probabilistic}.\footnote{
	For theoretical results on generative models based on scoring rule optimization, see \citet{PacchiardiEtAl2021}.
}
In our approach, a conditional generative model based on a neural network is trained by optimizing a suitable multivariate proper scoring rule which measures the discrepancy between the generated and true data, and replaces the GAN discriminator.
By conditioning the generative process on exogenous input variables, we enable the generative model to incorporate arbitrary input predictors beyond random noise only and to flexibly learn nonlinear relations to both the univariate forecast distributions and the multivariate dependence structure.

Using two case studies on spatial dependencies of temperature and wind speed forecasts at weather stations over Germany, we compare our conditional generative model with state-of-the-art two-step approaches to multivariate post-processing based on ECC and GCA.
For the univariate post-processing step, we consider models only based on ensemble predictions of the target variable as well as state-of-the-art neural network models with additional predictors.
This will allow for specifically investigating the benefits of incorporating additional information in the marginal distribution or the full multivariate forecast.

The remainder of the paper is organized as follows. Section \ref{data} introduces the datasets and Section \ref{standardmvpp} reviews the standard two-step post-processing methods.
Our conditional generative model is described in Section \ref{generative}, followed by the main results presented in Section \ref{results}.
Section \ref{conclusion} concludes with a discussion.
Python and R code with implementations of all methods is available online (\url{https://github.com/jieyu97/mvpp}), along with supplemental material with additional results (\url{https://github.com/jieyu97/mvpp/blob/main/CGM_supplement.pdf}).

\section{Data}\label{data}

\begin{figure}
	\centering
	\includegraphics[width=0.4\textwidth]{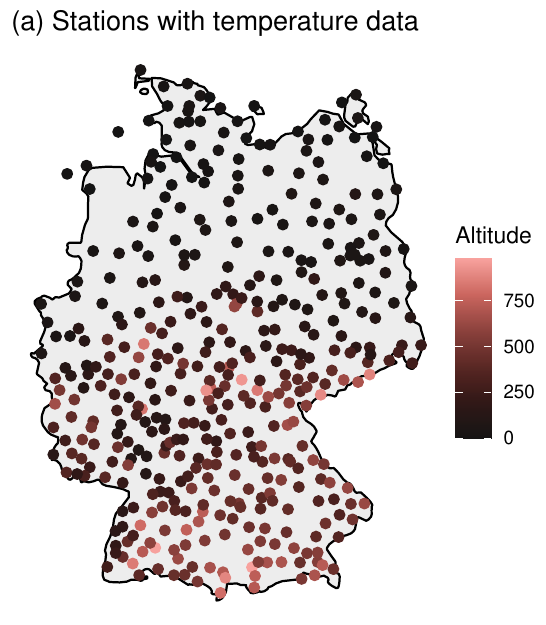}
	\includegraphics[width=0.4\textwidth]{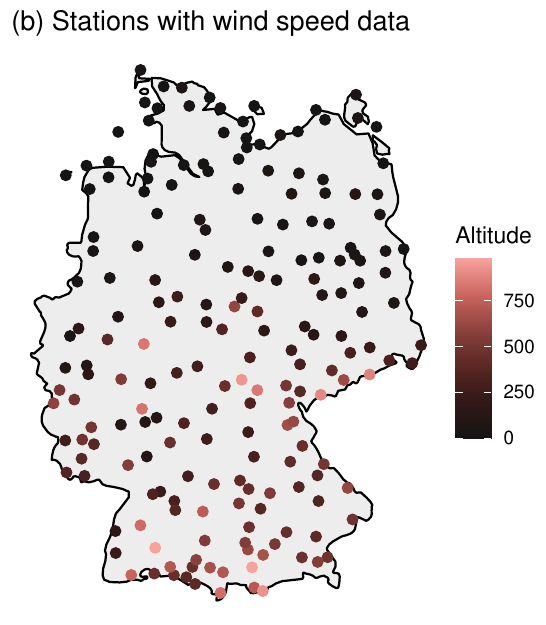}
	\caption{Locations of weather stations with (a) temperature and (b) wind speed observations.}
	\label{fig_location}
\end{figure}

Our study focuses on forecasts of 2-m temperature and 10-m wind speed with a forecast lead time of 48 hours. 
The dataset for temperature\footnote{
	The dataset is available from \url{https://doi.org/10.6084/m9.figshare.19453580}.
} is based on the one used in \cite{rasp2018neural}, while the dataset for wind speed\footnote{
	The dataset is available from \url{https://doi.org/10.6084/m9.figshare.19453622}. 
} was compiled for this study.
Both datasets are based on forecasts from the 50-member ensemble of the European Center of Medium-range Weather Forecasts (ECMWF) initialized at 00 UTC every day, which were obtained from the THORPEX Interactive Grand Global Ensemble (TIGGE) database \citep{TIGGE} on a $0.5^\circ \times 0.5^\circ$ grid over Europe.
Following the procedure outlined in \citet{rasp2018neural}, the ensemble forecasts of all meteorological predictor variables are interpolated to weather station locations over Germany. 
Stations with larger fractions of missing data and with altitudes above 1\,000 m are omitted to avoid outliers due to substantially different topographical properties when considering spatial dependencies. 
This results in a total of 419 stations in the temperature dataset and 198 stations in the wind speed dataset, the locations of which are shown in Figure \ref{fig_location}. Corresponding observations were obtained from the Climate Data Center\footnote{\url{https://www.dwd.de/DE/klimaumwelt/cdc/cdc_node.html}} of the German weather service.

In addition to ensemble forecasts of the target variables (temperature and wind speed, respectively), ensemble forecasts of several auxiliary predictor variables based on the selection in \citet{rasp2018neural} are available. Ensemble forecasts of all meteorological predictor variables are reduced to their mean and standard deviation. 
For the wind speed dataset, we also explicitly compute the wind speeds at different pressure levels from the corresponding wind components.
Further, we use a sine-transformed value of the day of the year
\footnote{
	The numerical value of the day of the year ranging from 1 to 366, $t$ is transformed via 
	$\texttt{doy} = \sin\left( \frac{t - 105}{366} \cdot 2\pi \right)$.
},
and relevant information about the station coordinates, altitudes, and orography (altitude of the model grid point) as additional input predictors for the post-processing models. Table \ref{tab_add_pred} provides an overview of all available predictors in both datasets.

\begin{table}
	\centering
	\caption{Available predictors for temperature and wind speed post-processing. The variables ws, ws\_pl850 and ws\_pl500 are derived from the corresponding U and V wind components and are not included in the temperature dataset. For wind speed forecasts, the orography (orog) information is missing.}
	\label{tab_add_pred}
	\begin{tabular}{ll}
		\toprule 
		Variable & Description \\
		\midrule
		\multicolumn{2}{l}{\textit{Meteorological variables}} \\
		\midrule
		t2m %*
		& 2-m temperature   \\
		d2m & 2-m dewpoint temperature   \\
		cape & Convective available potential energy    \\
		sp & Surface pressure    \\
		tcc & Total cloud cover    \\
		q\_pl850 & Specific humidity at 850 hPa    \\
		u\_pl850 & U component of wind at 850 hPa    \\
		v\_pl850 & V component of wind at 850 hPa    \\
		ws\_pl850 %*
		& Wind speed at 850 hPa  \\ 
		u\_pl500 & U component of wind at 500 hPa    \\
		v\_pl500 & V component of wind at 500 hPa    \\
		gh\_pl500 & Geopotential height at 500 hPa    \\
		ws\_pl500 %*
		& Wind speed at 500 hPa    \\
		u10 & 10-m U component of wind    \\
		v10 & 10-m V component of wind    \\
		ws %*
		& 10-m wind speed \\
		sshf & Sensible heat flux    \\
		slhf & Latent heat flux    \\
		ssr & Shortwave radiation flux    \\
		str & Longwave radiation flux  \\
		\midrule
		\multicolumn{2}{l}{\textit{Other predictors}} \\
		\midrule 
		lon & Longitude of station   \\
		lat & Latitude of station  \\
		alt & Altitude of station  \\
		orog %*
		& Altitude of model grid point  \\
		% \hline
		doy & Sine-transformed value of the day of the year  \\ 
		\bottomrule
	\end{tabular}
\end{table}

For both datasets, a total of 10 years of daily forecast and observation data from 2007--2016 are available. Following \citet{rasp2018neural}, we use data from 2007--2014 as training set and 2015 as validation set for choosing hyper-parameters and model specifications. Data from 2016 serves as out-of-sample test dataset and is not used during model training.

\section{Benchmark methods for multivariate ensemble post-processing}\label{standardmvpp}

This section introduces state-of-the-art approaches to multivariate post-processing which serve as benchmark methods for our generative machine learning method that will be introduced in Section \ref{generative}. 
We focus on the two-step approaches based on a combination of univariate post-processing models and copulas. In a first step, univariate post-processing is applied to ensemble forecasts for each margin (i.e., location, forecast horizon or target variable), and in a second step, the multivariate dependence structure is imposed upon the univariately post-processed forecast using a suitable copula. Sklar's theorem \citep{Sklar1959} provides the theoretical underpinning in that a multivariate CDF $H$ (our target) can be decomposed into a copula function $C$ modeling the dependence structures (this is what needs to be specified) and its marginal univariate CDFs $F_1,\ldots,F_D$ (here obtained via univariate post-processing) via
\[
H(z_1,\ldots,z_D)=C(F_1(z_1),\ldots,F_D(z_D))
\]
for $z_1,\ldots,z_D \in \mathbb{R}$.  

In the following, we first introduce methods for univariate post-processing and then discuss the use of copula functions for restoring multivariate dependencies. Our notation follows that of \citet{lerch2020simulation} and we denote the unprocessed $D$-dimensional ensemble forecasts of a weather variable with $M$ members by $\boldsymbol{X}_1, ..., \boldsymbol{X}_M \in\mathbb{R}^D$, where $\boldsymbol{X}_m = \left(X_m^{(1)}, ..., X_m^{(D)}\right)$ for $m=1,...,M$. 
The corresponding observation is $\boldsymbol{y} = \left(y^{(1)}, ..., y^{(D)}\right) \in\mathbb{R}^D$, and we use a generic index $d = 1,...,D$ to summarize the margins (in our case the locations when modeling spatial dependencies).

\subsection{Univariate post-processing}

Over the past years, a large variety of univariate post-processing methods has been proposed. In particular, the development of modern methods from ML for post-processing has been a focus of recent research interest. We refer to \citet{vannitsem2021statistical} for a general overview and to \citet{SchulzLerch2022} for a recent comparison in the context of wind gust forecasting. We restrict our attention to post-processing methods within the parametric distributional regression framework proposed by \citet{gneiting2005calibrated}. To specifically investigate the effect of improved marginal predictions, we consider a simple ensemble model output statistics approach \citep{gneiting2005calibrated} and a state-of-the-art method based on neural networks \citep{rasp2018neural} which allows for incorporating additional predictor information and flexibly model nonlinear relations to forecast distribution parameters. 

Within the distributional regression framework, the conditional probability distribution of the (univariate) variable of interest $y$, given ensemble predictions $X_1,...,X_M$  of the target variable, is modeled by a parametric distribution, $F_{\boldsymbol{\theta}}$, the parameters of which depend on the ensemble predictions via a link function $g$, i.e.,
\[
\boldsymbol{\theta} = g(X_1,...,X_M).
\]
Note that within the current subsection, we suppress the index of the dimension to simplify notation.

\subsubsection{Ensemble model output statistics (EMOS)}\label{emos}

Following \citet{gneiting2005calibrated}, the standard EMOS model for temperature (t2m) assumes a Gaussian predictive distribution
\[
y|X_1^{\text{t2m}},...,X_M^{\text{t2m}} \sim \mathcal{N}(\mu,\sigma)
\]
and uses ensemble forecasts of the target variable, $X_1^{\text{t2m}},...,X_M^{\text{t2m}}$, as sole predictors. The distribution parameters are linked to the ensemble mean and standard deviation (sd) via 
\begin{equation}\label{equ_params}
\mu = a_0 + a_1\cdot\text{mean}(X_1^{\text{t2m}},...,X_M^{\text{t2m}}) 
\quad \text{and} \quad 
\sigma = b_0 + b_1\cdot\text{sd}(X_1^{\text{t2m}},...,X_M^{\text{t2m}}).
\end{equation}

The EMOS coefficients $a_0, a_1, b_0, b_1$ are determined by minimizing the continuous ranked probability score (CRPS) on the training dataset, see Section \ref{scores} for details on verification metrics. We here implement local EMOS models by estimating separate sets of coefficients for different stations, which makes the models locally adaptive and typically leads to better performance compared with a global model that is jointly estimated for all stations, provided that a sufficient amount of training data is available \citep{LerchBaran2017}. 
While rolling training windows consisting of the most recent days only have often been used for the estimation of EMOS models, we use a static training period given by the entire training dataset, following common practice in the operational use of post-processing models and results from studies suggesting that the benefits of using long archives of training data often outweigh potential changes in the underlying NWP model or the meteorological conditions \citep{Lang2020}.

For wind speed, we proceed analogously, but follow \citet{thorarinsdottir2010probabilistic} and utilize a Gaussian distribution left-truncated at 0 as predictive distribution to ensure that no probability mass is assigned to negative wind speed values. Forecasts of wind speed (ws) are used as sole input predictors, i.e.,
\[
y|X_1^{\text{ws}},...,X_M^{\text{ws}} \sim \mathcal{N}_{[0,\infty)}(\mu,\sigma),
\]
where $\mathcal{N}_{[0,\infty)}(\mu,\sigma)$ denotes a truncated Gaussian distribution with cumulative distribution function
\[
F(z) = \Phi\left(\frac{\mu}{\sigma}\right)^{-1} \Phi\left(\frac{z-\mu}{\sigma}\right)
\]
for $z>0$ and 0 otherwise. By contrast to the situation for temperature, the choice of a parametric distribution is less clear for wind speed, and a large variety of distributions has been considered \citep{LerchThorarinsdottir2013,BaranLerch2015,ScheuererMoeller2015,BaranLerch2016,PantillonEtAl2018,baran2021truncated}. However, the differences across parametric distributions are usually only minor and are unlikely to effect our results and conclusions. To link the parameters of the truncated Gaussian distribution to the ensemble predictions, we proceed as in \eqref{equ_params}, i.e.,
\begin{equation}\label{equ_params2}
\mu = a_0 + a_1\cdot\text{mean}(X_1^{\text{ws}},...,X_M^{\text{ws}}) 
\quad \text{and} \quad 
\sigma = b_0 + b_1\cdot\text{sd}(X_1^{\text{ws}},...,X_M^{\text{ws}}).
\end{equation}

\subsubsection{Neural network methods for univariate post-processing}\label{nn}

Many standard approaches to post-processing, including the EMOS method introduced above, share a common limitation in that incorporating additional predictors beyond forecasts of the target variable is challenging. To do so, it would be necessary to manually specify the exact functional form of the dependencies between the distribution parameters and all available input predictors in equations \eqref{equ_params} and \eqref{equ_params2}. 
Over the past years, a variety of ML methods have been developed to address this issue \citep{vannitsem2021statistical}.  
\cite{rasp2018neural} propose a neural network (NN) approach, where the distribution parameters are obtained as the output of a NN which allows for learning arbitrary nonlinear relations between input predictors and distribution parameters in an automated, data-driven manner. We refer to this approach as the distributional regression network (DRN).

In our implementations, we follow \cite{rasp2018neural} and use a NN with one hidden layer. All available predictors (listed in \tabref{tab_add_pred}) except for the date information are normalized to the range $[0,1]$ using a min-max scaler and are then used as inputs to the NN which returns the distribution parameters $\mu$ and $\sigma$ as outputs. 
A single NN model is estimated jointly for all stations, using the CRPS as a custom loss function. 
The model predictions are made locally adaptive by the use of embeddings of the station identifiers, a technique that was originally proposed in natural language processing \citep{pennington2014glove}.
Our model architecture and implementation choices directly follow those of \cite{rasp2018neural} for temperature, where we employ a Gaussian predictive distribution. For wind speed, we use a truncated Gaussian predictive distribution as in the corresponding EMOS model, and apply a softplus activation, 
\begin{equation*}
\mathrm{softplus}(z) = \log(1 + \exp(z)),
\end{equation*}
to the output layer to ensure positivity of the distribution parameters which helps to avoid numerical issues. 

The results presented in \cite{rasp2018neural} indicate that the DRN approach to post-processing leads to substantial improvements over state-of-the-art benchmark methods, and subsequent research has generalized the methodology to other target variables and characterizations of the forecast distribution \citep{bremnes2020ensemble,ScheuererEtAl2020,Chapman2022,SchulzLerch2022}.
For our purposes here it will be interesting to investigate whether the improvements in the univariate predictive performance directly extends to improved multivariate predictions when coupling the univariate DRN models with the re-ordering techniques introduced below. 

\subsection{Multivariate extensions using copulas}\label{copula}

Univariate post-processing methods are intended to correct systematic errors in the marginal distributions. However, multivariate dependencies are lost when univariate post-processing is applied separately for each margin (e.g., each station in our application), and need to be restored. A variety of methods for restoring multivariate dependencies via copula functions have been proposed over the past years. We here limit our discussion to the popular ensemble copula coupling and Gaussian copula approach, and refer to \citet{schefzik2013uncertainty, wilks2015multivariate} and \citet{lerch2020simulation} for overviews and comparisons. In our descriptions, we follow \citet{lerch2020simulation} and refer to their Section 2 for further mathematical details and references.

\subsubsection{Ensemble copula coupling}\label{ecc}

Given univariately post-processed marginal distributions, a sample of the same size as the raw ensemble, $M$, is drawn from each predictive marginal distribution. 
While several sampling schemes have been proposed \citep{schefzik2013uncertainty,hu2016stratified}, we only consider the use of equidistant quantiles at levels $\frac{1}{M+1}, ..., \frac{M}{M+1}$.
Ensemble copula coupling (ECC) is based on the assumption that the ensemble forecasts are informative about the true multivariate dependence structure, and it makes use of the rank order structure of the raw ensemble member forecasts to rearrange the sampled values, with possible ties resolved at random. This can be interpreted as a non-parametric, empirical copula approach, which we refer to as ECC-Q.

A widely used alternative non-parametric approach is the Schaake shuffle \citep{clark2004schaake} which proceeds as ECC, but reorders the sampled quantiles based on the rank order structure of past observations instead of the ensemble forecasts.
As noted in the introduction, comparative studies have often found similar predictive performances between the Schaake shuffle and ECC \citep[e.g.,][]{LakatosEtAl2022}.
For the datasets at hand, initial tests indicated a slightly worse performance compared to ECC-Q (not shown), and we thus only use ECC-Q as a non-parametric benchmark approach to retain focus.

\subsubsection{Gaussian copula approach}\label{gca}

In contrast to ECC-Q, the Gaussian copula approach \citep[GCA;][]{PinsonGirard2012,moller2013multivariate} is based on a parametric Gaussian copula.
In a first step of the application of GCA, a set of past observations is transformed into latent standard Gaussian observations via 
\[
\tilde{y}^{(d)} = \Phi^{-1}\left( F_\theta^{(d)} \left( y^{(d)}\right) \right)
\]
for all dimensions $d = 1,...,D$, where $ F_\theta^{(d)}$ denotes the corresponding forecast distributions obtained via univariate post-processing.
In a next step, multivariate random samples $\boldsymbol{Z}_1,...,\boldsymbol{Z}_M$ are randomly drawn from a $D$-dimensional Gaussian distribution $\mathcal{N}^{(D)}(\boldsymbol{0},\Sigma)$ with a mean vector of 0 and an empirical correlation matrix $\Sigma$ based on the observations transformed into a latent Gaussian space in the first step.
The final post-processed GCA forecasts are then obtained via 
\[
{X}^{\text{GCA}\,(d)}_{m} = \left( F_\theta^{(d)} \right)^{-1} \left( \Phi \left( Z_m^{(d)} \right) \right) 
\]
for $m = 1,...,M$ and $d = 1,...,D$.

In addition to the assumption of a parametric copula, the main difference of GCA to ECC-Q is given by the use of past observations to determine the dependence template. 
While the number of GCA ensemble members is not limited by the size of the raw ensemble, we here only consider $M = 50$ to ensure comparability across methods.

To summarize, in the following we will consider four two-step approaches based on the available combinations of methods for univariate post-processing (EMOS, DRN) and copula-based modeling of multivariate dependencies (ECC-Q, GCA) for both target variables (temperature and wind speed) which will serve as benchmarks for our generative ML approach. These benchmark methods will be abbreviated by EMOS+ECC, EMOS+GCA, DRN+ECC and DRN+GCA, respectively.

\section{Generative models for multivariate distributional regression}\label{generative}

We propose a novel approach to multivariate ensemble post-processing using generative machine learning techniques. 
Moving beyond the previous two-step strategy of separately modeling marginal distributions and multivariate dependence structure, this new class of data-driven multivariate distributional regression models allows for obtaining multivariate probabilistic forecasts directly as output of a NN.
The proposed generative models provide a non-parametric way to generate post-processed multivariate samples without any distributional assumptions on the marginal distributions or the multivariate dependencies.
Further, by allowing for incorporating additional predictors beyond forecasts of the target variable and for generating an arbitrary number of samples, the proposed generative models address key limitations of state-of-the-art two-step approaches to multivariate post-processing.

\subsection{Deep generative models}

Implicit generative models aim to provide a representation of the probability distribution of a target variable by defining a stochastic procedure to generate samples $\boldsymbol{Y}_1, \dots, \boldsymbol{Y}_{\nout}$ from the distribution of interest \citep{mohamed2016learning}.
The only input to the generative model 
\begin{equation}
\boldsymbol{Y}_i = g_{\boldsymbol{\theta}}(\boldsymbol{Z}_i)
\end{equation}
are samples $\boldsymbol{Z}_1, \dots \boldsymbol{Z}_{\nout}$ from a simple base distribution, e.g., a standard multivariate normal distribution $\boldsymbol{Z} \sim N(\boldsymbol{0}, \boldsymbol{I})$.
The learnable map $g$ is typically parameterized by a deep NN with parameters $\boldsymbol{\theta}$.

As implicit generative models do not provide a tractable density of the target distribution, classic  parameter estimation procedures such as maximum likelihood estimation are infeasible.
Generative adversarial networks \citep{goodfellow2014generative} sidestep this problem by specifying an additional classification model, the discriminator, which is trained to discriminate between the true data and the generated samples.
The generator model is then trained to maximize the misclassification rate of the discriminator.
In the context of ensemble post-processing, \citet{DaiHemri2021} propose a GAN-based model for generating spatially coherent maps of total cloud cover forecasts.

While impressive results have been achieved by GANs, in particular in the context of image processing and computer vision, the adversarial training process is often considered to be complex and unstable \citep{Gui2022}. 
Therefore, several alternative approaches for training generative models have been developed.
From a statistical perspective, generative moment matching networks \citep[GMMNs;][]{li2015generative, dziugaite2015training} are a particularly interesting class of generative models.
GMMNs replace the discriminator with the maximum mean discrepancy \citep[MMD;][]{gretton2012kernel}, a kernel-based two-sample test statistic which can be used to measure distances on the space of probability distributions.
The training objective is then to minimize a Monte Carlo estimate of the MMD.
However, GMMNs only approximate the unconditional data distribution $P(\boldsymbol{Y})$ while we are interested in the conditional distribution $P(\boldsymbol{Y}|\boldsymbol{X})$, i.e., we aim for a conditional generative model of the form
\begin{equation}
\boldsymbol{Y}_i = g_{\boldsymbol{\theta}}(\boldsymbol{X}, \boldsymbol{Z}_i).
\end{equation}
Our approach builds on recent work on probabilistic model averaging in energy forecasting \citep{janke2020probabilistic} and uses the energy score \citep[ES;][see Section \ref{scores} for details]{gneiting2007strictly} as loss function to train a conditional generative model.
The ES is a multivariate strictly proper scoring rule which is derived from the energy distance \citep{szekely2003statistics}, which in turn is a special case of the MMD \citep{sejdinovic2013equivalence}.
Training a conditional generative model based on ES optimization allows for generating multivariate post-processed forecasts that simultaneously correct systematic biases and dispersion errors as well as systematic errors in the multivariate dependence structure.

\subsection{Notation}

We now introduce the notation that will be used for describing the model architecture and training process. As input predictors at every weather station location $d=1,...,D$, we have NWP forecasts of $K$ different weather variables available (see Table \ref{tab_add_pred}), each in the form of an ensemble of size $M$, i.e., $X_{k,m}^{(d)}$ is the value of the $m$th ensemble member for variable $a_k$ at location $d$, with $m = 1,...,M$, $k = 1,...,K$, and $d = 1,...,D$. 
The ensemble weather forecasts are reduced to the corresponding ensemble mean 
$\mu(\boldsymbol{X}_k^{(d)})= \frac{1}{M}\sum_m X_{k,m}^{(d)}$ and standard deviation $\sigma(\boldsymbol{X}_k^{(d)})$. 
In the following, we will use 
\[ 
\bar{\boldsymbol{X}}^{(d)} = \left[\mu\left(\boldsymbol{X}_1^{(d)}\right), \dots, \mu\left(\boldsymbol{X}_K^{(d)}\right)\right]^T
\quad
\text{and}
\quad
\boldsymbol{s}^{(d)} = \left[\sigma\left(\boldsymbol{X}_1^{(d)}\right), \dots, \sigma\left(\boldsymbol{X}_K^{(d)}\right)\right]^T
\]
to denote vectors of size $K$ which contain the mean ensemble predictions of all variables at location $d$ and the corresponding standard deviations, respectively.
As additional static input predictors, we use the vector $\textbf{loc}^{(d)} = [\text{lat}^{(d)}, \text{lon}^{(d)}, \text{alt}^{(d)}, \text{orog}^{(d)}]^T$ with location-specific information, as well as the (scalar) sine-encoded day of the year, $\text{doy}$.
Further, we will denote a single sample from the $D_\text{latent}$-dimensional noise distribution by $\boldsymbol{z}_i=[z_1, \dots, z_{D_\text{latent}}]^T \sim \mathcal{N}^{D_\text{latent}}(\bm{0},\bm{I})$,
% i.e., $z_j \sim \mathcal{N}(0,1)$ for $j = 1, ..., D_\text{latent}$, 
and denote a single final output sample from the $D$-dimensional target distribution by $\hat{\boldsymbol{y}}_i, i = 1,...,\nout$.

\subsection{Model architecture and training}

\begin{figure}
	\centering
	\includegraphics[width=\textwidth]{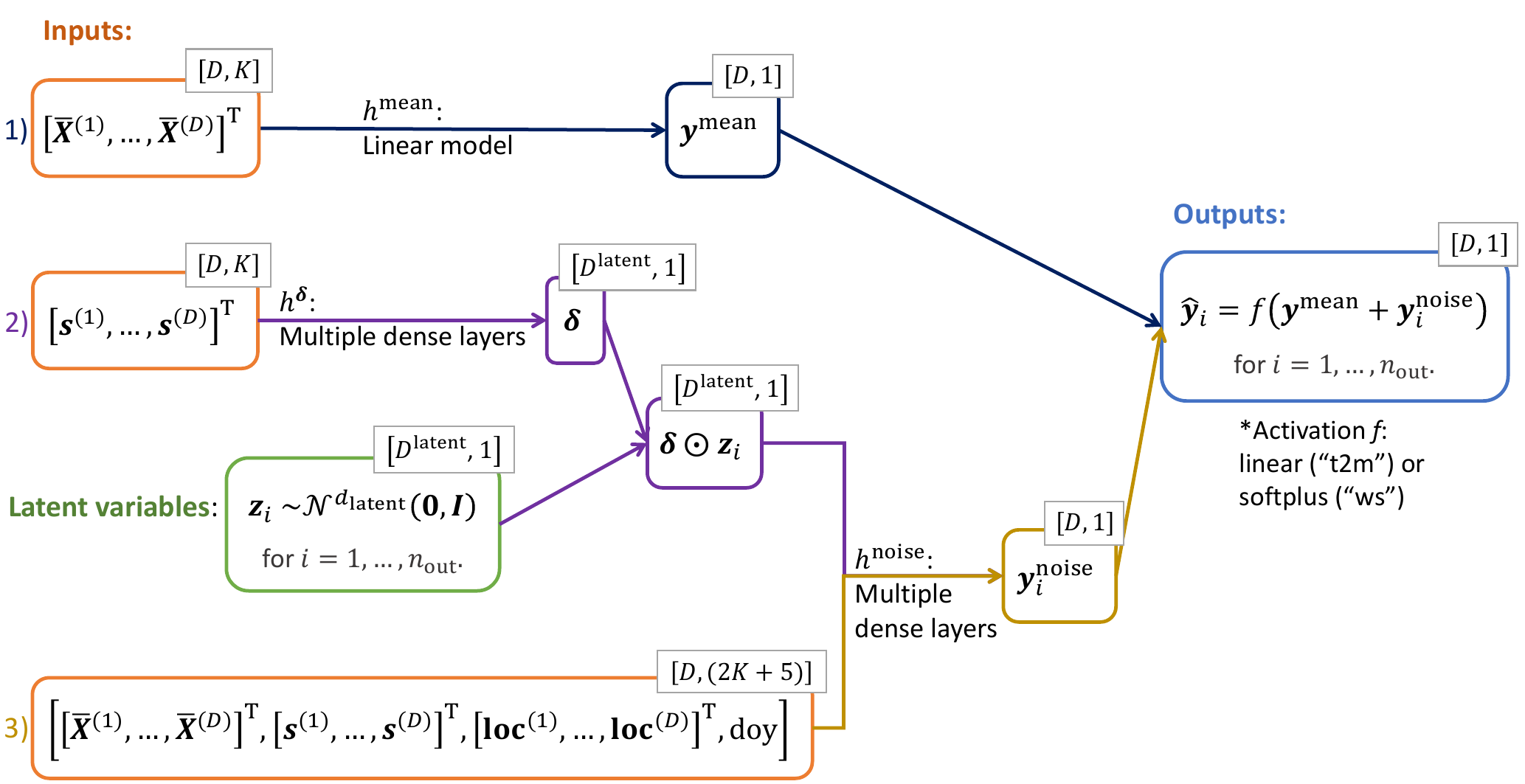}
	\caption{Schematic illustration of the conditional generative model. The dimensions of the tensors at each step are indicated in the small box.}
	\label{fig_cgm}
\end{figure}

A schematic overview of our conditional generative model (CGM) is provided in Figure \ref{fig_cgm}. The same basic model structure is used for both temperature and wind speed prediction, and we will highlight relevant differences in the following. 
The final output of the model is given by $D$-dimensional multivariate samples drawn from the post-processed joint distribution, 
$\{\hat{\boldsymbol{y}}_i, i = 1, ..., \nout\}$,
which are composed of a mean component $\boldsymbol{y}^{\text{mean}} \in\mathbb{R}^D$, and a noise component $\boldsymbol{y}_{i}^{\text{noise}}\in\mathbb{R}^D$ that depends on the sample $\boldsymbol{z}_i \in \mathbb{R}^{D_\text{latent}}$ from the latent noise distribution for $i = 1,...,\nout$. 

Our CGM architecture allows for incorporating arbitrary input predictors and for specifically tailoring the model structure to incorporate relevant exogenous information in the different components of the target distribution by separating the mean and noise component. 
To efficiently propagate the uncertainty inherent to the NWP forecasts, we dynamically reparametrize the latent distributions of the generative model conditionally on the standard deviations of the NWP ensemble forecasts.
The overall model consists of three modules with different sets of inputs, which will be described in the following. 

The first part of the model aims to learn the multivariate mean component $\boldsymbol{y}^{\text{mean}}$ of the forecast distribution and can be considered as a multivariate deterministic bias-correction step of the mean ensemble forecasts, conditional on the available additional predictors. 
This mean module thus maps  mean ensemble forecasts of all weather input variables, including the target variables, to $\boldsymbol{y}^{\text{mean}}$, i.e.,
\begin{equation*}
\boldsymbol{y}^{\text{mean}} = h^\text{mean}\left(\bar{\boldsymbol{X}}^{(1)}, ..., \bar{\boldsymbol{X}}^{(D)}\right).
\end{equation*}

We utilize a linear model without hidden layers for $h^\text{mean}$, i.e., we have for each dimension $d = 1,...,D$
\begin{equation*}
y^{\text{mean}\,(d)} = w_{0,d} +  \sum_{k=1}^K w_{k,d} \cdot \mu\left(\boldsymbol{X}^{(d)}_{k}\right).
\end{equation*}

The remaining two parts of the model aim to learn the noise component $\boldsymbol{y}^{\text{noise}}$ of the multivariate forecast distribution conditional on the available predictor information.
The generative structure of our CGM approach becomes apparent in the second part of the model, where $\nout$ random samples of $D_\text{latent}$-dimensional latent variables are drawn independently from  a standard multivariate normal distribution, i.e.,
\[
\boldsymbol{z}_1,...,\boldsymbol{z}_{\nout} \sim \mathcal{N}^{D_\text{latent}}(\bm{0},\bm{I}),
% F(\theta),
\]
where $\boldsymbol{z}_i\in\mathbb{R}^{D^\text{latent}}$ for $i = 1,...,\nout$. 
The number of samples $\nout$ generated in this step directly controls the final number of output samples obtained from the multivariate CGM forecast distribution and thus allows for generating an arbitrary number of post-processed samples.\footnote{Note that the description in the following will focus on a single sample $\boldsymbol{z}_i$. During the model training process described in more detail below, we generate $\nout$ independent CGM predictions.}
In a next step, the noise encoder module aims to encode the uncertainty from the ensemble weather forecasts into the latent distribution by adjusting the scale of the latent variables via a linear mapping.
To that end, we first model the conditional variance of the latent distribution via a fully connected NN
\[
\boldsymbol{\delta} = h^{\boldsymbol{\delta}} \left( \boldsymbol{s}^{(1)}, ..., \boldsymbol{s}^{(D)} \right),
\]
where an exponential activation function is applied in the output layer to ensure positivity of the variances $\boldsymbol{\delta} \in \mathbb{R}^{D_\text{latent}}$.
The scaling coefficient vector $\boldsymbol{\delta} = [\delta_1, ..., \delta_{D_\text{latent}}]^T$ is then used to adjust the variance of the latent noise variables, i.e.,
\begin{equation*}
\tilde{\boldsymbol{z}_i} = \boldsymbol{\delta} \odot \boldsymbol{z}_i
\end{equation*}
for $i = 1,...,\nout$.\footnote{Note that we now have $\tilde{\boldsymbol{z}}_i \sim N(\boldsymbol{0}, diag(\boldsymbol{\delta}))$, i.e., it is an isotropic Gaussian with a variance conditional on the ensemble standard deviations.
}

In the final part of the model, the scale-adjusted latent variables from the noise encoder module are now combined with additional predictors to yield the final conditional noise component $\boldsymbol{y}^{\text{noise}}_i$. This noise decoder module is a fully connected NN
\[
\boldsymbol{y}_i^\text{noise} = h^\text{noise}\left(\bar{\boldsymbol{X}}^{(1)}, ..., \bar{\boldsymbol{X}}^{(D)}, \boldsymbol{s}^{(1)}, ..., \boldsymbol{s}^{(D)}, \textbf{loc}^{(1)}, ..., \textbf{loc}^{(D)}, \text{doy}, \tilde{\boldsymbol{z}}_i \right).
\]

The final output of the model is given by the collection of realizations from the multivariate forecast distribution 
\begin{equation*}
\hat{\boldsymbol{y}}_i = \boldsymbol{y}^{\text{mean}} + \boldsymbol{y}_i^{\text{noise}},
\end{equation*}
for $i = 1,...,\nout$ for temperature. For wind speed, we additionally apply a softplus activation function 
\begin{equation}
\hat{\boldsymbol{y}}_i =  \log \big(1 + \exp(\boldsymbol{y}^{\text{mean}} + \boldsymbol{y}_i^{\text{noise}}) \big)
\end{equation}
to ensure  non-negativity of the obtained wind speed forecasts. 

During training, for each training example $(\boldsymbol{X}_n, \boldsymbol{y}_n)$, where $\bm{X}_n$ and $\bm{y}_n$ represent the inputs and true observations at forecast case $n = 1,...,N$, respectively, we generate $\nout$ independent predictions $\hat{\boldsymbol{y}}_1, \dots, \hat{\boldsymbol{y}}_{\nout}$ by querying the model $\nout$ times. Each time the model generates predictions from a different sampled noise vector $\boldsymbol{z}_i, i = 1,...,\nout$, but uses the same inputs $\boldsymbol{X}_n$. 
The training procedure is formalized in Algorithm \ref{alg:TrainingAlgo}.\footnote{Note that this procedure might be slow on a CPU but the innermost loop in Algorithm \ref{alg:TrainingAlgo} is fully parallelizable and can thus be efficiently implemented on GPUs.}

\begin{algorithm}
	\small
	\SetKwInOut{Input}{Input}
	\SetKwInOut{Output}{Output}
	\Input{data $\{(\boldsymbol{X},\boldsymbol{y})_n\}_{n=1}^N$, initial model parameters $\boldsymbol{\theta}_0$, number of samples $n_{\text{train}}$, number of batches $B$, learning rate $\eta$}
	\Output{Model parameters $\boldsymbol{\theta}^*$}
	\For{$N_{epochs}$}{
		Get mini batch $\{(\boldsymbol{X},\boldsymbol{y})_{b}\}_{b=1}^{B}$\;
		For each sample $b$ generate a set of $n_{\text{train}}$ random noise samples $\{[\boldsymbol{z}_b^1, \ldots, \boldsymbol{z}_b^{n_{\text{train}}}]\}_{b=1}^{B}$ \\
		\For{$b=1, \ldots, B$}{
			\For{$s=1, \ldots, n_{\text{train}}$}{
				Compute forward pass $\hat{\boldsymbol{y}}_b^s \leftarrow g_{\boldsymbol{\theta}}(\boldsymbol{X}_b, \boldsymbol{z}_b^s)$
			}
		}
		Compute loss over batch $L \leftarrow \frac{1}{B}\sum_{b} \text{ES} \big(\boldsymbol{y}_b, [\hat{\boldsymbol{y}}^1_b, \dots, \hat{\boldsymbol{y}}_b^{n_{\text{train}}}] \big)$\;
		Compute gradient $\nabla_{\boldsymbol{\theta}} L$\;
		Update learning rate $\eta$ using ADAM\;
		Update model parameters $\boldsymbol{\theta} \leftarrow \boldsymbol{\theta} -  \eta \nabla_{\boldsymbol{\theta}}L$		
	}
	\caption{CGM training algorithm}
	\label{alg:TrainingAlgo}
\end{algorithm}

The CGM parameters, i.e., the weights and biases of $h^\text{mean}$, $h^{\boldsymbol{\delta}}$ and $ h^\text{noise}$, are estimated by optimizing the energy score (see Section \ref{scores}) as a loss function tailored to the specific situation of multivariate probabilistic forecasting based on an implicit representation of the forecast distribution in the form of a sample of size $n_{\text{train}} = 50$.
We follow common practice in the machine learning literature and generate an ensemble of CGMs by repeating the estimation process multiple times from different random initializations to account for the randomness of the training process based on stochastic gradient descent methods \citep{Lakshminarayanan2017, SchulzLerch2022ens}. 
Unless indicated otherwise, we will generate a set of $\nout = 50$ multivariate samples to ensure comparability with the benchmark methods when making predictions on the test set, and do so by repeating the model estimation 10 times and generating 5 samples each.

\subsection{Implementation details and hyper-parameter choices}

Multiple hyper-parameters need to be determined for the CGM implementation. For the specific setup of the individual modules of the model ($h^\text{mean}$, $h^{\boldsymbol{\delta}}$ and $ h^\text{noise}$), we use two hidden layers with 100 nodes each in the noise decoder module $h^\text{noise}$ for both target variables. As for the noise encoder module $h^{\bm{\delta}}$, we use two hidden layers with 100 nodes each for wind speed but one linear dense layer for temperature. An elu activation function is applied in all hidden layers. Initial experiments indicated no substantial changes in the resulting model performance, we thus do not further optimize this component of the model architecture. Note that we also tested a nonlinear model based on a fully connected NN for $\boldsymbol{y}^{\text{mean}}$ which did not lead to substantial improvements. Details are provided in the supplemental material.

The number of latent variables, $D^\text{latent}$, the learning rate and the batch size are initially determined using the distributed asynchronous hyper-parameter optimization technique from \cite{bergstra2013making} implemented in the \texttt{hyperopt} package.
Based on the automated hyper-parameter optimization and selected additional experiments we set $D^\text{latent}$ to 10, and use a learning rate of 0.001 and a batch size of 64 for both target variables. The results are relatively robust to changes in these hyper-parameters. Exemplary results are illustrated in the form of ablation studies in the supplemental material.
The model is trained using stochastic gradient descent optimization based on the Adam optimizer \citep{kingma2014adam}, where we employ an early stopping criterion with a patience of 10 epochs to avoid overfitting. 
The maximum number of epochs is set to 300 but generally, the CGM training and validation losses converge after a few epochs.

All available predictors (listed in \tabref{tab_add_pred}) except for the target variable and date information are normalized by removing the mean and scaling to unit variance using the standard scaler from the \texttt{scikit-learn} package \citep{scikit-learn}. 

\section{Results}\label{results}

We here compare the CGM predictions with the various benchmark methods based on two-step procedures of separately modeling marginal distributions and multivariate dependencies introduced in Section \ref{copula}, i.e., EMOS+ECC, EMOS+GCA, DRN+ECC and DRN+GCA. To do so, we first provide some background information on forecast evaluation methods (\secref{scores}), and present the general setup of our experiments (\secref{tests}) and univariate results (\secref{sec:univ}). 
The main focus is on the multivariate forecast performance presented in Section \ref{sec:multiv}. Finally, the effect of the number of samples generated from the multivariate forecast distribution is investigated in Section \ref{sec:cgm-sample-size}.

\subsection{Forecast evaluation methods}\label{scores}

We briefly review key concepts relevant to the evaluation of probabilistic forecasts and refer to \citet[Appendix A]{rasp2018neural} and \citet[Appendix B]{lerch2020simulation} for details. 
The general goal of probabilistic forecasting is to maximize the sharpness of a predictive distribution subject to calibration \citep{gneiting2007probabilistic}. 
In order to assess calibration and sharpness simultaneously, proper scoring rules are now widely used for comparative evaluation of probabilistic forecasts.
A scoring rule $S(F,y)$ assigns a numerical score to a pair of a predictive distribution $F\in\mathcal{F}$ and a realizing observation $y\in\Omega$, where $\mathcal{F}$ is a class of probability distributions on $\Omega$. It is called proper if the true distribution of the observation minimizes the expected score, i.e., if $\mathbb{E} S(G,Y) \leq \mathbb{E} S(F,Y)$ if $Y\sim G$ for all $F,G\in\mathcal{F}$ \citep{gneiting2007strictly}.
The continuous ranked probability score \citep[CRPS;][]{matheson1976scoring} given by 
\begin{equation*}
\mathrm{CRPS}(F,y) = -\int_{-\infty}^\infty \big( F(z) - \one\{z \geq y\} \big)^2 dz,
\end{equation*}
where $\one$ denotes the indicator function and $F$ is assumed to have a finite first moment, is a popular proper scoring rule for univariate probabilistic forecasts (i.e., $\Omega \subset \mathbb{R}$). 
Closed-form analytical expressions are available for many parametric forecast distributions as well as probabilistic forecasts given in the form of a simulated sample \citep{JordanEtAl2019}. 

While the definition of proper scoring rules can in principle be straightforwardly extended towards multivariate settings with $\Omega\subset\mathbb{R}^D$, many practical questions remain open and a variety of multivariate proper scoring rules have been proposed over the past years \citep{PetropoulosEtAl2022}. Most of these multivariate proper scoring rules focus on multivariate probabilistic forecasts in the form of samples from the forecast distributions.

With the notation introduced in Section \ref{standardmvpp}, the energy score \citep[ES;][]{gneiting2007strictly},
\[
\text{ES}(F,\bm{y}) = \frac{1}{M}\sum_{i=1}^M \| \boldsymbol{X}_i - \boldsymbol{y} \| - \frac{1}{2M^2} \sum_{i = 1}^M\sum_{j = 1}^M \| \boldsymbol{X}_i - \boldsymbol{X}_j \|,
\]
where $\|\cdot\|$ is the Euclidean norm on $\mathbb{R}^D$, and the variogram score of order $p$ \citep[VS$^p$;][]{scheuerer2015variogram},
\[
\text{VS}^p(F,\bm{y}) = \sum_{i=1}^D\sum_{j=1}^D w_{i,j} \left( \left|y^{(i)} - y^{(j)} \right|^p - \frac{1}{M}\sum_{k=1}^M \left|X_k^{(i)} - X_k^{(j)} \right|^p \right)^2.
\]
are the most popular examples of multivariate proper scoring rules.
In the definition of the VS, $w_{i,j} \geq 0$ is a non-negative weight for pairs of component combinations and $p$ is the order of the VS. We use an unweighted (i.e. $w_{i,j}=1$ for all $i,j$) version of the VS with order $p = 0.5$ throughout, following suggestions of \cite{scheuerer2015variogram} and utilizing implementations provided in \citet{JordanEtAl2019}.
Other multivariate proper scoring rules have been proposed including copula scores focusing on the dependence structure \citep{ZielBerk2019} and weighted versions of ES and VS \citep{AllenEtAl2022}, and we present additional results for some of these scores in the supplemental material.

To compare forecasting methods based on a proper scoring rule with respect to a benchmark, we will often calculate the associated skill score. With the mean score of the forecasting method of interest over a test dataset, $\bar S_{\text{f}}$, the corresponding mean scores of the benchmark, $\bar S_{\text{ref}}$, and the (typically hypothetical) optimal forecast, $\bar S_{\text{opt}}$, the skill score $SS_{\text{f}}$ is calculated via
\begin{eqnarray*}
	SS_{\text{f}} = \dfrac{ \bar S_{\text{ref}} - \bar S_{\text{f}} }{ \bar S_{\text{ref}} - \bar S_{\text{opt}}}.
\end{eqnarray*}
For the scoring rules considered below, $S_{\text{opt}} = 0$. Skill scores are positively oriented with a maximum value of 1, values of 0 indicating no improvement over the benchmark and negative values indicating a worse predictive performance than the benchmark. 

We further apply Diebold-Mariano (DM) tests of equal predictive performance \citep{Diebold1995} to assess the statistical significance of score differences between multivariate post-processing methods. To compare two forecasting methods $F$ and $G$ with based on a scoring rule $S$ and corresponding mean scores $\bar S^F_n = \frac{1}{n} \sum_{i=1}^n S(F_i,\boldsymbol{y}_i)$ and $\bar S^G_n = \frac{1}{n} \sum_{i=1}^n S(G_i,\boldsymbol{y}_i)$ over $n$ forecast cases, we employ two-sided tests based on the DM test statistic
\begin{eqnarray*}
	t_n = \sqrt{n} \dfrac{\overline{S}^F_n - \overline{S}^Q_n}{\hat{\sigma}_n}, \quad \text{where} \quad 
	\hat{\sigma}_n = \dfrac{1}{n} \sum_{i = 1}^n\left( S (F_i, \boldsymbol{y}_i) - S (G_i, \boldsymbol{y}_i) \right)^2.
\end{eqnarray*}
Under standard regularity conditions, $t_n$ is asymptotically standard normal under the null hypothesis of equal predictive performance. 
We use the DM tests with a nominal level of $\alpha = 0.05$ and apply a Benjamini–Hochberg procedure \citep{Benjamini1995} to account for multiple testing, see \citet{SchulzLerch2022} for details.

\subsection{Setup of the multivariate post-processing experiments}\label{tests}

To evaluate the multivariate forecast performance of different post-processing methods, we repeatedly sub-sample the station datasets described in Section \ref{data}. Focusing on spatial dependencies over geographically close stations in a setting that aims to mimic practical applications, we fix a number of dimensions $D \in \{5,10,20\}$, and proceed as follows.

We randomly pick a station and then select the $(D - 1)$ stations which are geographically closest to obtain a set of $D$ stations, based on which we implement the multivariate post-processing methods as described in Sections \ref{standardmvpp} and \ref{generative}.
Next, we apply the scoring rules introduced above to obtain corresponding mean scores over the test set, i.e., data from the calendar year 2016, and compute the corresponding skill scores and DM test statistics.

To account for uncertainties, the above procedure is repeated 100 times for both temperature and wind speed. For skill score computations, EMOS+ECC is used as a reference method throughout and for all methods, we generate 50 multivariate samples from all post-processed forecast distributions to ensure consistency and allow for a fair comparison.

\subsection{Univariate results}\label{sec:univ}

While the focus of our study is on multivariate post-processing, the univariate predictive performance constitutes an important component of the overall forecast quality. Here, we therefore first focus on the univariate, marginal predictions of different post-processing methods to investigate the performance of our CGM approach in this setting. 

\begin{figure}
	\centering
	\includegraphics[width=0.6\textwidth]{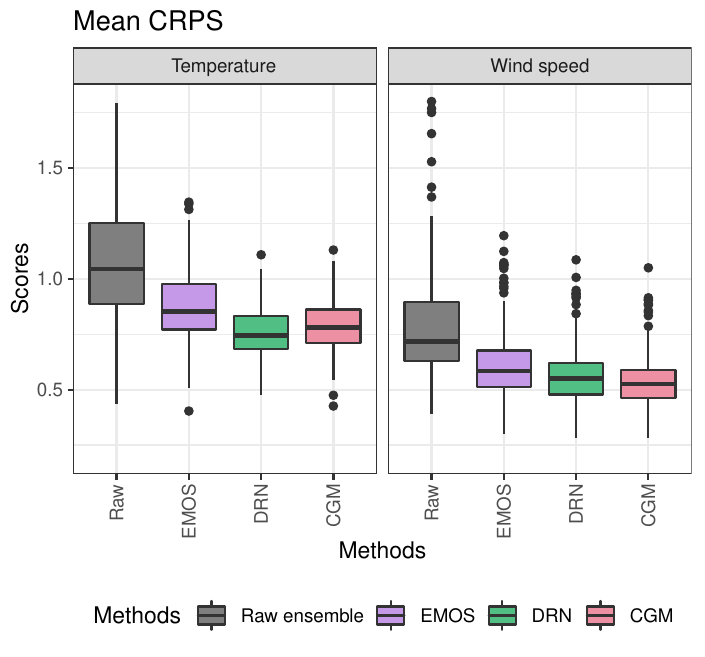}
	\caption{Boxplots of mean CRPS values of different multivariate post-processing methods with $D=5$, including the scores of raw ensemble forecasts. The scores are based on 242 unique stations in case of temperature, and 178 unique stations in case of wind speed.}
	\label{fig_dim5_crps}
\end{figure}

To evaluate the univariate forecast performance in the experimental setup described above, we restrict our attention to the experiments for $D=5$ and compute the mean CRPS over unique sets of stations present in the 100 repetitions of the sub-sampled station datasets. In case a station occurs multiple times in the randomly selected sets of stations, we only use data from the first occurrence within the 100 repetitions.
We compare the univariate performance of our CGM to the raw ensemble forecasts and two univariate post-processing approaches discussed earlier, i.e., EMOS and DRN, and Figure \ref{fig_dim5_crps} shows boxplots of the mean CRPS values over the corresponding unique sets of stations for temperature and wind speed. 
As expected, all univariate post-processing methods notably improve the forecast performance over the raw ensemble predictions, and the variability among different stations is reduced.
For temperature, the CGM forecasts generally outperform the EMOS predictions and are slightly worse than the DRN post-processed forecasts, where the mean CRPS over all stations is improved from 0.89 for EMOS to 0.76 for DRN and 0.79 for CGM. 
For wind speed, the CGM provides the overall best forecasts and clearly outperforms the DRN approach, with a mean CRPS improved from 0.62 for EMOS and 0.58 for DRN to 0.54 for CGM.
Additional results are provided in the supplemental material, including an assessment of calibration which indicates that all post-processing methods generally provide relatively well-calibrated forecasts.

\subsection{Multivariate results}\label{sec:multiv}

% rounded to 2 digits
\begin{table}
	\caption{Mean multivariate scores of different multivariate post-processing methods for temperature and wind speed, averaged over the 100 repetitions of the simulation experiment.  \label{tab_mean_scores}}
	\centering
	\begin{tabular}{l@{\hskip 0.45cm}lc@{\hskip 0.45cm}C{1.45cm}C{1.45cm}C{1.45cm}C{1.45cm}C{1.45cm}C{1.45cm}}
		\toprule
		Variable & Score & $D$ & Raw ens. & EMOS+ ECC & EMOS+ GCA & DRN+ ECC & DRN+ GCA & CGM \\
		\midrule
		\multirow{7}{*}{Temperature} & \multirow{3}{*}{ES} & 5 & 2.81 & 2.27 & 2.27 & \textbf{1.97} & \textbf{1.97} & \textbf{1.97} \\
		& & 10 & 4.22 & 3.37 & 3.37 & 2.91 & \textbf{2.90} & 2.91 \\
		& & 20 & 6.09 & 4.87 & 4.87 & \textbf{4.21} & 4.22 & 4.26 \\
		\cmidrule{2-9}
		& \multirow{3}{*}{VS} & 5 & 8.22 & 4.81 & 4.36 & 4.12 & 3.74 & \textbf{3.50} \\
		& & 10 & 39.0 & 22.6 & 21.0 & 19.5 & 18.0 & \textbf{16.9} \\
		& & 20 & 153 & 96.7 & 92.8 & 85.0 & 80.7 & \textbf{77.8} \\
		\midrule
		\multirow{7}{*}{Wind speed} & \multirow{3}{*}{ES} & 5 & 2.44 & 1.69 & 1.68 & 1.56 & 1.55 & \textbf{1.44} \\
		& & 10 & 3.67 & 2.55 & 2.53 & 2.31 & 2.30 & \textbf{2.16} \\
		& & 20 & 5.04 & 3.52 & 3.51 & 3.23 & 3.22 & \textbf{3.04} \\
		\cmidrule{2-9}
		& \multirow{3}{*}{VS} & 5 & 9.49 & 4.37 & 4.00 & 4.01 & 3.66 & \textbf{3.31} \\
		& & 10 & 39.7 & 20.2 & 19.0 & 18.0 & 16.9 & \textbf{15.4} \\
		& & 20 & 153 & 82.5 & 78.9 & 75.6 & 72.3 & \textbf{67.0} \\
		\bottomrule
	\end{tabular}
\end{table}

\begin{figure}[h]
	\centering
	\includegraphics[width=\textwidth]{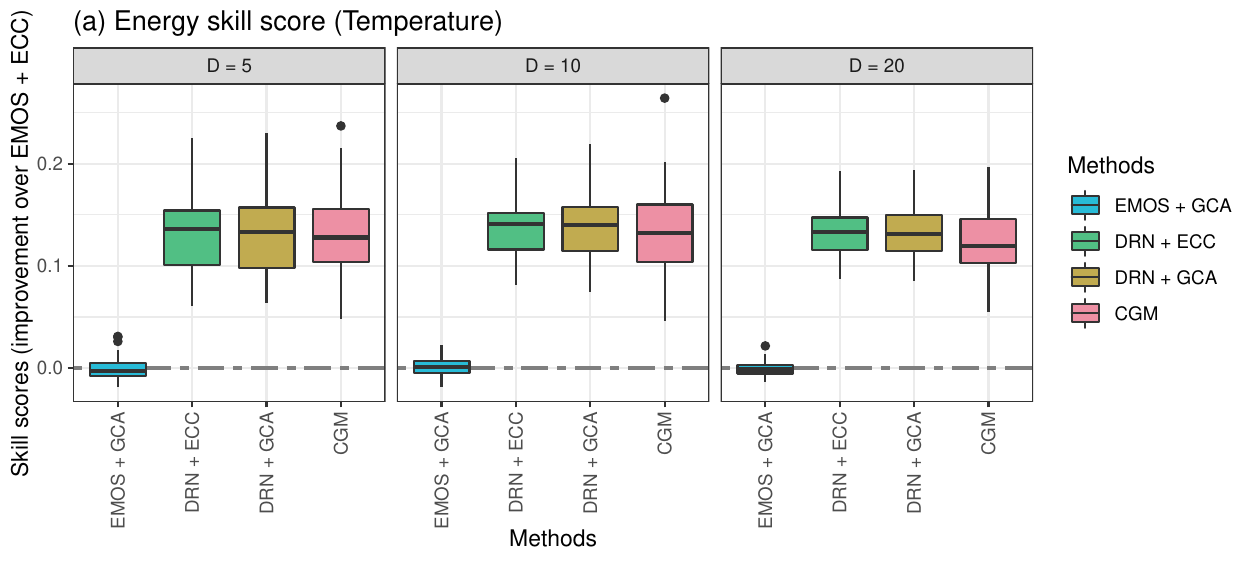}
	\includegraphics[width=\textwidth]{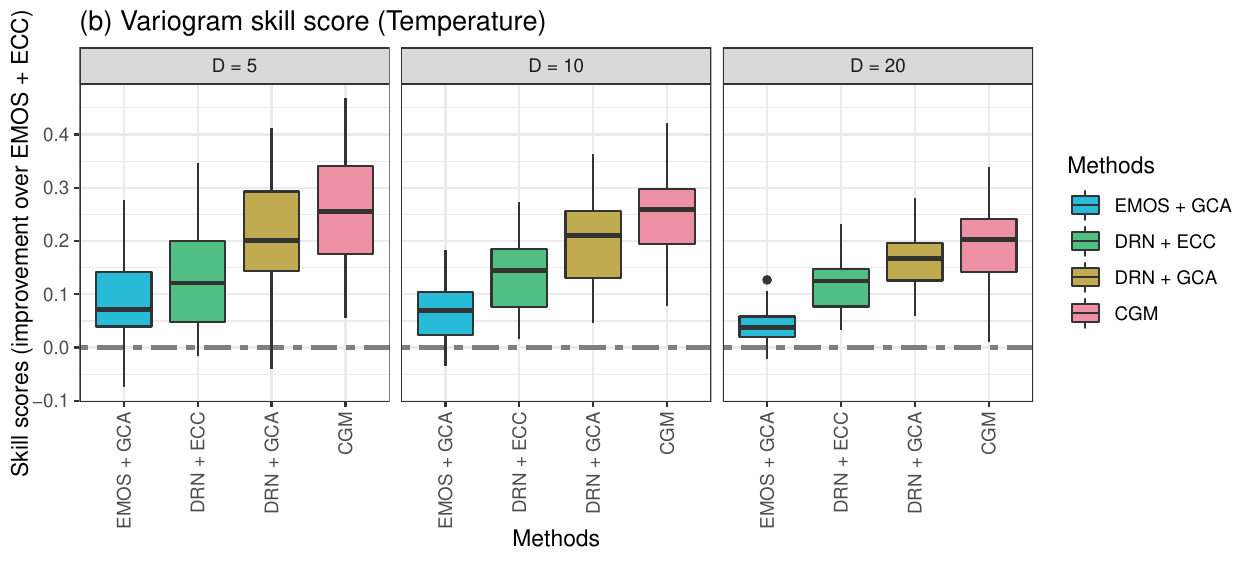}
	\caption{Boxplots of (a) energy skill scores and (b) variogram skill scores of different multivariate post-processing methods for temperature across the 100 repetitions of the experiment with different sets of stations. EMOS+ECC is used as reference forecast in both cases.}
	\label{fig_tem_mvscores}
\end{figure}

We now turn to the key part of our results and compare the multivariate performance of our CGM approach to the two-step post-processing approaches used as benchmark methods. 
Table~\ref{tab_mean_scores} summarizes the mean scores of different multivariate post-processing models for temperature and wind speed. For both target variables, all post-processing methods clearly improve the raw ensemble predictions. In comparison to the EMOS-based models, the DRN-based models show clear improvements in the multivariate performance. Regarding the choice of the reordering method, ECC and GCA lead to similar results in terms of the ES, but the GCA-based forecasts lead to better performance in terms of the VS. The CGM consistently provides the best multivariate forecasts and outperforms the state-of-the-art approaches across the variables, dimensions and evaluation metrics. The only exception to this observation are temperature forecasts evaluated with the ES, where the DRN+ECC and DRN+GCA models provide slightly better forecasts for $D=10$ and $D=20$. The values of all considered multivariate scoring rules increase with the spatial dimension $D$, which is to be expected from the definition of the scoring rules and consistent with findings in the extant literature.

\begin{figure}
	\centering
	\includegraphics[width=\textwidth]{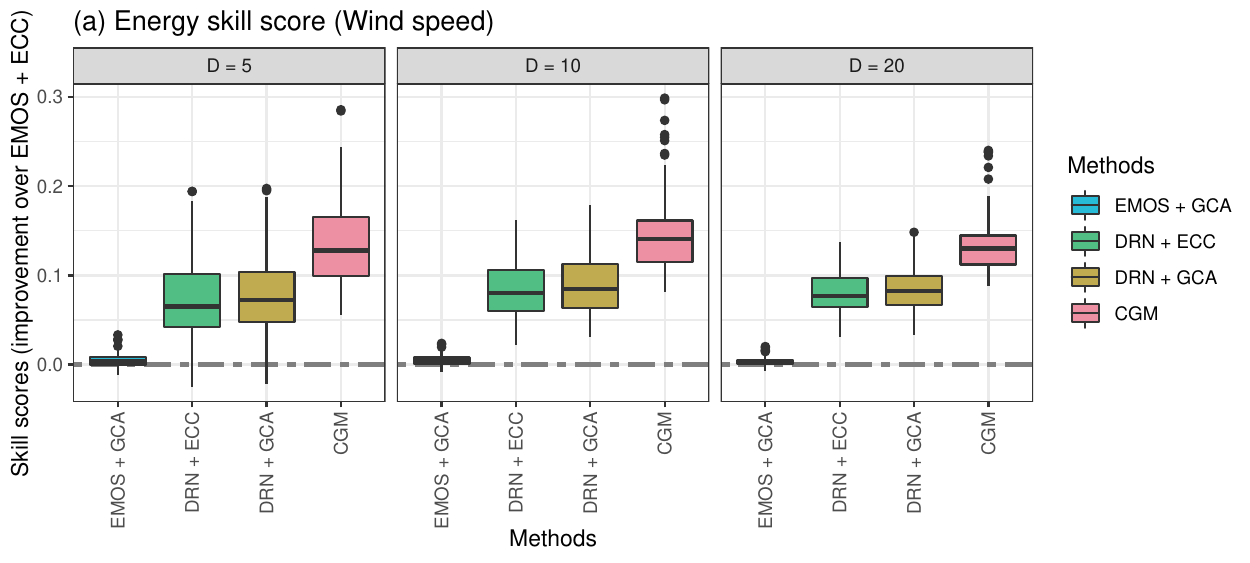}
	\includegraphics[width=\textwidth]{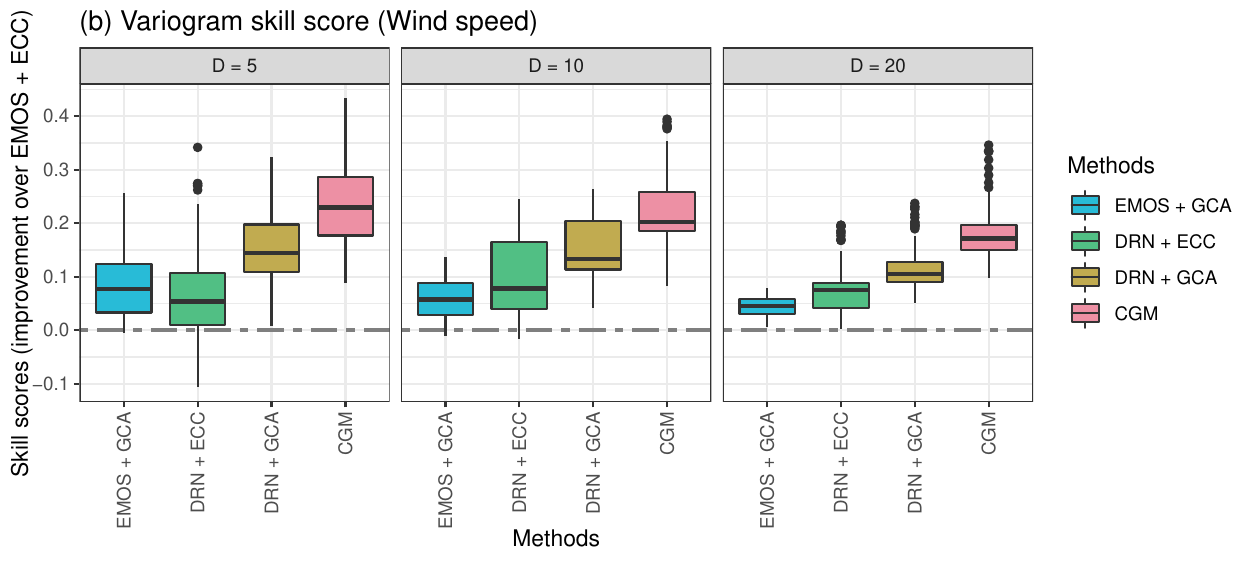}
	\caption{As \figref{fig_tem_mvscores}, but for wind speed.}
	\label{fig_ws_mvscores}
\end{figure}

To investigate the variability across the selected sets of stations, Figures \ref{fig_tem_mvscores} and \ref{fig_ws_mvscores} show boxplots of the multivariate skill scores for temperature and wind speed, respectively, using EMOS+ECC as reference method. For the temperature forecasts (Figure \ref{fig_tem_mvscores}), the DRN-based two-step methods and the CGM provide consistent and comparable improvements over the reference in terms of the ES. The relative improvements in terms of the VS show a larger variability across the sets of stations, and indicate a superior performance of the CGM forecasts. Among the considered two-step approaches, applying GCA leads to improvements over ECC in terms of the VS, but similar results in terms of the ES. The above observations apply to all considered spatial dimensions and we do not observe any obvious trends in terms of $D$, indicating that consistent improvements can be observed also in the higher-dimensional settings in the experiments.

Qualitatively similar results can be observed for the multivariate wind speed forecasts shown in Figure \ref{fig_ws_mvscores}. The main difference to the results for temperature are the notably larger improvements of the CGM forecasts in comparison to the DRN-based approaches, particularly in terms of the ES. Interestingly, in terms of the VS at $D=5$, the DRN+ECC models here fail to outperform the EMOS+GCA forecasts despite the incorporation of additional predictor variables in the marginal distributions. A potential explanation for this observation is that the disadvantages due to the misspecifcations in the multivariate dependence structure of the raw ensemble forecasts which serve as a dependence template for ECC outweigh the benefits of incorporating additional predictors in the marginal distributions. Similar to the temperature forecasts, the DRN+GCA model results in better forecasts than the DRN+ECC approach, but performs notably worse than the CGM.

Tables \ref{tab_dmtable_10dim_t2m} and \ref{tab_dmtable_10dim_ws} show the rejection rates of DM tests of equal predictive performance and thus allow for quantifying the statistical significance of the observed score differences between the multivariate post-processing methods across the repetitions of the experiment. 
In line with the results from above, we find that the observed score differences are significant to a large degree.
The CGM forecasts show significant improvements over the other methods with the exception of temperature forecasts evaluated with the ES. For example, the null hypothesis of equal predictive performance is rejected in at least 99\% of all cases in favor of the CGM forecasts when compared to all two-step approaches for wind speed, where we further do not observe a single case where the CGM is outperformed significantly by any one of the other models. In particular in terms of the ES, the differences between the multivariate re-ordering approaches tend to not be significant, with the notable exception of the EMOS+GCA approach showing comparable performance to the DRN+ECC model in case of wind speed forecasts evaluated with the VS.

Additional verification results including assessments of multivariate calibration and results for other multivariate proper scoring rules are available in the Supplemental Material.

\begin{table}[p]
	\centering 
	\caption{Proportion of pair-wise Diebold-Mariano tests for the temperature forecasts indicating statistically significant ES or VS differences after applying a Benjamini-Hochberg procedure to account for multiple testing for a nominal level of 0.05 of the corresponding one-sided tests. The $(i,j)$-entry in the $i$-th row and $j$-th column indicates the proportion of tests where the null hypothesis of equal predictive performance of the corresponding one-sided DM test is rejected in favor of the model in the $i$-th row when compared to the model in the $j$-th column. The remainder of the sum of $(i,j)$- and $(j,i)$-entry to 1 is the proportion of tests where the score differences are not significant. We consider only the case $D = 10$ here, additional results are available in the supplemental material.}
	\label{tab_dmtable_10dim_t2m}
	% \small 
	\begin{tabular}{lC{1.45cm}C{1.45cm}C{1.45cm}C{1.45cm}C{1.45cm}}
		\toprule
		\multicolumn{6}{c}{Energy score} \\
		\midrule 
		& EMOS+ ECC & EMOS+ GCA & DRN+ ECC & DRN+ GCA & CGM  \\
		\midrule
		EMOS+ECC & & \cellcolor{red!10}0.04 & \cellcolor{red!5}0.00 & \cellcolor{red!5}0.00 & \cellcolor{red!5}0.00  \\
		EMOS+GCA & \cellcolor{red!15}0.17 & & \cellcolor{red!5}0.00 & \cellcolor{red!5}0.00 & \cellcolor{red!5}0.00  \\
		DRN+ECC & \cellcolor{red!60}1.00 & \cellcolor{red!60}1.00 & & \cellcolor{red!10}0.03 & \cellcolor{red!10}0.08  \\
		DRN+GCA & \cellcolor{red!60}1.00 & \cellcolor{red!60}1.00 & \cellcolor{red!15}0.16 & & \cellcolor{red!10}0.02  \\
		CGM & \cellcolor{red!55}0.99 & \cellcolor{red!60}1.00 & \cellcolor{red!15}0.13 & \cellcolor{red!10}0.07 & \\
		\midrule 
		\multicolumn{6}{c}{Variogram score} \\
		\midrule 
		& EMOS+ ECC & EMOS+ GCA & DRN+ ECC & DRN+ GCA & CGM  \\
		\midrule
		EMOS+ECC & & \cellcolor{red!5}0.00 & \cellcolor{red!5}0.00 & \cellcolor{red!5}0.00 & \cellcolor{red!5}0.00    \\
		EMOS+GCA & \cellcolor{red!45}0.75 & & \cellcolor{red!5}0.00 & \cellcolor{red!5}0.00 & \cellcolor{red!5}0.00   \\
		DRN+ECC & \cellcolor{red!55}0.98 & \cellcolor{red!45}0.75 & & \cellcolor{red!5}0.00 & \cellcolor{red!5}0.00   \\
		DRN+GCA & \cellcolor{red!60}1.00 & \cellcolor{red!60}1.00 & \cellcolor{red!55}0.95 & & \cellcolor{red!5}0.00   \\
		CGM & \cellcolor{red!60}1.00 & \cellcolor{red!60}1.00 & \cellcolor{red!55}0.97 & \cellcolor{red!50}0.84 & \\
		\bottomrule
	\end{tabular}
	\bigbreak 
	\caption{As Table \ref{tab_dmtable_10dim_t2m}, but for wind speed.}
	\label{tab_dmtable_10dim_ws}
	% \small 
	\begin{tabular}{lC{1.45cm}C{1.45cm}C{1.45cm}C{1.45cm}C{1.45cm}}
		\toprule
		\multicolumn{6}{c}{Energy score} \\
		\midrule 
		& EMOS+ ECC & EMOS+ GCA & DRN+ ECC & DRN+ GCA & CGM  \\
		\midrule
		EMOS+ECC & & \cellcolor{red!5}0.00 & \cellcolor{red!5}0.00 & \cellcolor{red!5}0.00 & \cellcolor{red!5}0.00  \\
		EMOS+GCA & \cellcolor{red!20}0.26 & & \cellcolor{red!5}0.00 & \cellcolor{red!5}0.00 & \cellcolor{red!5}0.00  \\
		DRN+ECC & \cellcolor{red!55}0.99 & \cellcolor{red!55}0.99 & & 0.01 & \cellcolor{red!5}0.00  \\
		DRN+GCA & \cellcolor{red!60}1.00 & \cellcolor{red!60}1.00 & \cellcolor{red!20}0.21 & & \cellcolor{red!5}0.00  \\
		CGM & \cellcolor{red!60}1.00 & \cellcolor{red!60}1.00 & \cellcolor{red!60}1.00 & \cellcolor{red!55}0.99 & \\
		\midrule 
		\multicolumn{6}{c}{Variogram score} \\
		\midrule 
		& EMOS+ ECC & EMOS+ GCA & DRN+ ECC & DRN+ GCA & CGM  \\
		\midrule
		EMOS+ECC & & \cellcolor{red!5}0.00 & \cellcolor{red!5}0.00 & \cellcolor{red!5}0.00 & \cellcolor{red!5}0.00   \\
		EMOS+GCA & \cellcolor{red!50}0.89 & & \cellcolor{red!25}0.33 & \cellcolor{red!5}0.00 & \cellcolor{red!5}0.00 \\
		DRN+ECC & \cellcolor{red!50}0.84 & \cellcolor{red!30}0.45 & & \cellcolor{red!5}0.00 & \cellcolor{red!5}0.00 \\
		DRN+GCA & \cellcolor{red!60}1.00 & \cellcolor{red!55}0.91 & \cellcolor{red!50}0.88 & & \cellcolor{red!5}0.00   \\
		CGM & \cellcolor{red!60}1.00 & \cellcolor{red!60}1.00 & \cellcolor{red!60}1.00 & \cellcolor{red!55}0.99 & \\
		\bottomrule
	\end{tabular}
\end{table}

\subsection{CGM sample size}\label{sec:cgm-sample-size}

\begin{figure}[h]
	\centering
	\includegraphics[width=0.49\textwidth]{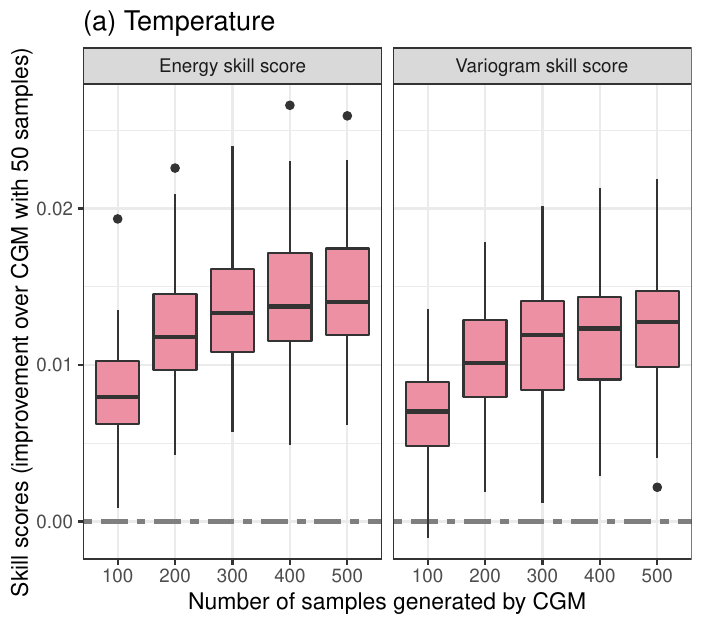}
	\includegraphics[width=0.49\textwidth]{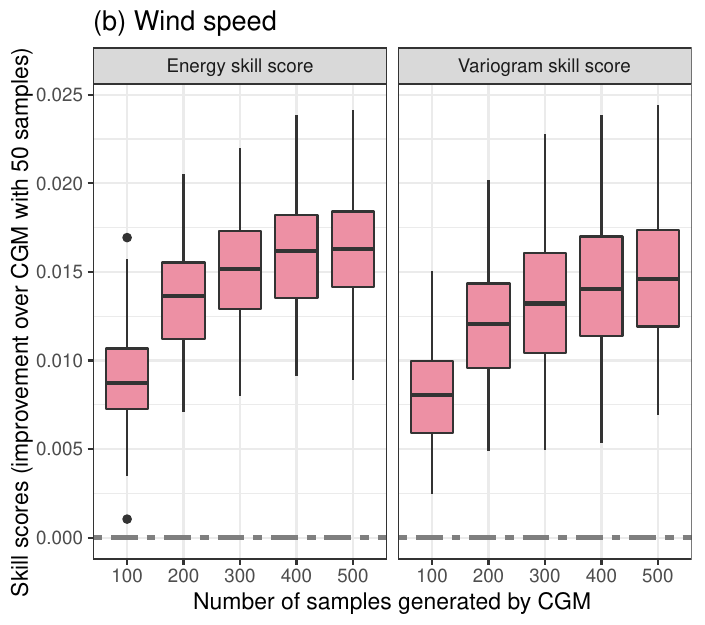}
	\caption{Boxplots of energy skill scores and variogram skill scores of CGM forecasts with different numbers of samples generated from the multivariate post-processed forecast distribution for (a) temperature and (b) wind speed over the 100 repetitions of the experiment with different sets of stations. The CGM approach with $\nout = 50$ is used as reference forecast and we only consider the case $D = 10$ here.}
	\label{fig_cgm500_dim20_mvscores}
\end{figure}

In addition to incorporating arbitrary predictor variables, a particular advantage of the proposed CGM approach over ECC is that the generative procedure allows for generating an arbitrary number of samples from the multivariate predictive distribution instead of being limited to the number of ensemble members. To investigate the effect of the size $\nout$ of the generated CGM ensemble, Figure \ref{fig_cgm500_dim20_mvscores} shows boxplots of the multivariate skill scores as functions of the ensemble size, based on the 100 repetitions of the experiment for $D = 10$. As before, we repeat the model estimation procedure of the CGM approach 10 times and generate $\frac{\nout}{10}$ samples each time to obtain a final post-processed ensemble of size $\nout$.

Compared to the reference setting of 50 CGM ensemble members, generating a larger sample from the post-processed distributions generally improves the predictive performance, with median improvements in terms of the energy and variogram score of up to around 1.5\%. The median skill score values increase notably up to an ensemble size of 200, after which some minor improvements can be observed. Additional results on Diebold-Mariano tests and other values of $D$ are provided in the supplemental material.

Given a fixed CGM ensemble size $\nout$, various ensembling strategies for obtaining these $\nout$ forecasts could be devised. For example, to obtain 50 CGM members, one could repeat the CGM model estimation 50 (or 25, 10, 5, 2, 1) time(s) and generate 1 (or 2, 5, 10, 25, 50) sample(s) each, respectively. While we found that in general, increasing the number of model runs leads to larger improvements in predictive performance compared to increasing the number of generated samples, this needs to be balanced against the added computational costs for repeating the CGM estimation. More details on the effects of different ensembling strategies are provided in Appendix \ref{appendix:ensembling}.

\section{Discussion and conclusions}\label{conclusion}

We propose a nonparametric multivariate post-processing method based on a conditional generative machine learning model which is trained by optimizing a suitable multivariate proper scoring rule. In our CGM approach, an arbitrary number of samples from the multivariate forecast distribution is directly obtained as output of a generative deep neural network which allows for incorporating arbitrary input predictors beyond ensemble predictions of the target variables only. By circumventing the two-step structure of the state-of-the-art multivariate post-processing approaches, the generative model aims to simultaneously correct systematic errors in the marginal distributions and the multivariate dependence structure. 
By contrast to the standard two-step methods, our CGM approach  does not require the choice of parametric models. 
Further, our CGM architecture can be specifically tailored to incorporate relevant exogenous information and domain knowledge in the different components of the target distribution. For example, our noise encoder module allows for dynamically reparametrizing the latent distributions of the generative model conditionally on the standard deviations of the NWP ensemble forecasts to efficiently propagate uncertainty information.

In two case studies on spatial dependencies of temperature and wind speed forecasts at weather stations over Germany, our generative model outperforms state-of-the-art two-step methods for multivariate post-processing where univariate post-processing via DRN models is combined with ECC and GCA. Our CGM approach provides improvements in terms of the univariate forecast performance at individual stations, and produces the best overall multivariate forecasts in terms of the energy score and the variogram score. The observed score differences are statistically significant for a large fraction of the random repetitions of the experiments, even when compared to the best-performing benchmark methods. Overall, there are no clear differences in the performance across the considered spatial dimensions of 5, 10 and 20 stations, indicating that the CGM approach works well also in higher-dimensional settings. Regarding the two target variables, we observed more pronounced improvements over the state-of-the-art two-step methods for wind speed, potentially mainly due to larger improvements in the univariate forecast performance. In terms of the two considered multivariate proper scoring rules, the relative improvements are generally larger in terms of the variogram score, indicating that our CGM approach particularly succeeds in better modeling the multivariate dependence structure. The only case where we did not observe notable differences to the performance of the benchmark methods were the results for temperature in terms of the energy score. 

The clear improvements in terms of the predictive performance are likely due to key conceptual advantages of our conditional generative models over the two-step approaches, in particular their ability to incorporate arbitrary predictor variables in both the modeling of the marginal distributions and the multivariate dependencies. CGM architectures without additional predictors beyond ensemble forecasts of the target variables reached a better multivariate predictive performance than EMOS+ECC and EMOS+GCA models, but failed to outperform the DRN-based models. Details are provided in Appendix \ref{appendix:inputs}.

Two minor disadvantages of the CGM approach are on the one hand given by the slightly increased variability across the random repetitions of the experiments due to the generative procedure, which can lead to single outliers with a worse predictive performance. Generating ensembles of CGM predictions can help to alleviate this, in particular with an increased number of sub-ensembles and sample size \citep{SchulzLerch2022ens}, see Appendix \ref{appendix:ensembling}. 
On the other hand, the CGM approach is conceptually somewhat simpler than the two-step methods in that it does not require any parametric assumptions on univariate forecast distributions or multivariate dependencies and produces forecasts in a single step only, however, the computational costs of model training are larger. That said, the computational costs of CGM for multivariate post-processing are still negligible compared to the computational costs of generating the raw ensemble forecasts and will not be a limiting factor in research or operations. For example, for a fixed set of $D = 20$ stations, the process of estimating an ensemble of 10 CGMs and generating 5 samples each takes around 2 minutes on a Nvidia RTX A5000 GPU.

Over the past years, many techniques have been developed in order to better interpret and understand what machine learning methods have learned, in particular for NNs \citep[see][for an overview from a meteorological perspective]{mcgovern2019making}.
Methods from interpretable machine learning have been applied in the literature on univariate post-processing \citep{TaillardatEtAl2016,rasp2018neural,SchulzLerch2022}, but the problem is more involved for multivariate post-processing, in particular for the generative models proposed here. 
While there has been some progress for GANs \citep{chen2016infogan,adel2018discovering}, interpretation is challenging for generative models 
\citep{Zhou2022} and the application of standard methods such as permutation feature importance is not straightforward.

Our results provide several avenues for further generalization and analysis. While we have focused on spatial dependencies across observation stations, it would be interesting to investigate the performance of the CGM approach on a gridded dataset. 
Motivated by the potential of score-based generative models to achieve comparable performance to GANs in image generation tasks \citep{song2020improved}, applications to multivariate post-processing similar to the GAN models proposed in \citet{DaiHemri2021} constitute a natural starting point for future work.
In addition, while the focus of our case studies was on the multivariate forecast performance, the results presented in Section \ref{sec:univ} indicate that the univariate CGM forecasts show competitive performance even with state-of-the-art NN-based post-processing models applied for the marginal distributions, despite being trained in a multivariate setting. Therefore, it would also be interesting to investigate the potential of the generative models for univariate probabilistic forecasting in more detail, ideally in conjunction with considering theoretical aspects such as the effects of choosing different proper scoring rules for optimization \citep{PacchiardiEtAl2021}. 
Further, the CGM architecture could be combined with additional predictors that aim to incorporate information from flow-dependent large-scale spatial structures in the raw forecast fields. For example, \citet{LerchPolsterer2022} propose convolutional autoencoder NNs to learn low-dimensional representations of spatial forecast fields which could be used as additional CGM inputs to achieve a spatially-informed modeling of multivariate dependencies.
Finally, as an alternative two-step strategy for multivariate post-processing, it is also possible to employ generative ML methods to learn
conditional copula functions \citep{JankeEtAl2021} in the second step which allow for incorporating arbitrary additional predictors. 

The evaluation of multivariate probabilistic forecasts continues to represent an important methodological challenge, despite relevant recent work on multivariate proper scoring rules \citep{ZielBerk2019,alexander2022evaluating,AllenEtAl2022}. 
Regarding the evaluation of the CGM forecasts, the question on how to best differentiate between the contributions of improvements in univariate and multivariate components of the overall forecast performance measured via multivariate proper scoring rules is a particular challenge. 
Another important aspect is the evaluation of multivariate extreme events \citep{LerchEtAl2017}, where recent work from \citet{AllenEtAl2022} could serve as a starting point for systematically investigating the effect of the sample size of our CGM and alternative approaches on the ability of the post-processing models to provide reliable and accurate multivariate predictions of extreme events.

\section*{Acknowledgments}

The research leading to these results has been done within the Young Investigator Group ``Artificial Intelligence for Probabilistic Weather Forecasting'' funded by the Vector Stiftung. In addition, this project has received funding from the KIT Center for Mathematics in Sciences, Engineering and Economics under the seed funding programme. We thank Nina Horat, Benedikt Schulz and Tilmann Gneiting for helpful comments and discussions, and Sam Allen for providing code for the weighted multivariate scoring rules.

\bibliographystyle{myims2}
\bibliography{reference}

\begin{thebibliography}{85}
\expandafter\ifx\csname natexlab\endcsname\relax\def\natexlab#1{#1}\fi
\expandafter\ifx\csname url\endcsname\relax
  \def\url#1{\texttt{#1}}\fi
\expandafter\ifx\csname urlprefix\endcsname\relax\def\urlprefix{URL }\fi
\providecommand{\eprint}[2][]{\url{#2}}

\bibitem[{Adel et~al.(2018)Adel, Ghahramani and Weller}]{adel2018discovering}
{Adel, T.}, {Ghahramani, Z.} and {Weller, A.} (2018).
\newblock Discovering interpretable representations for both deep generative
  and discriminative models.
\newblock In \textit{Proceedings of the 35th International Conference on
  Machine Learning}, vol.~80. PMLR, 50--59.

\bibitem[{Alexander et~al.(2022)Alexander, Coulon, Han and
  Meng}]{alexander2022evaluating}
{Alexander, C.}, {Coulon, M.}, {Han, Y.} and {Meng, X.} (2022).
\newblock Evaluating the discrimination ability of proper multi-variate scoring
  rules.
\newblock \textit{Annals of Operations Research} 1--27.

\bibitem[{Allen et~al.(2022)Allen, Ginsbourger and Ziegel}]{AllenEtAl2022}
{Allen, S.}, {Ginsbourger, D.} and {Ziegel, J.} (2022).
\newblock Evaluating forecasts for high-impact events using transformed kernel
  scores.
\newblock Preprint, \urlprefix\url{https://arxiv.org/abs/2202.12732}.

\bibitem[{Baran and Lerch(2015)}]{BaranLerch2015}
{Baran, S.} and {Lerch, S.} (2015).
\newblock Log-normal distribution based {E}nsemble {M}odel {O}utput
  {S}tatistics models for probabilistic wind-speed forecasting.
\newblock \textit{Quarterly Journal of the Royal Meteorological Society},
  {141}, 2289--2299.

\bibitem[{Baran and Lerch(2016)}]{BaranLerch2016}
{Baran, S.} and {Lerch, S.} (2016).
\newblock Mixture {EMOS} model for calibrating ensemble forecasts of wind
  speed.
\newblock \textit{Environmetrics}, {27}, 116--130.

\bibitem[{Baran and M{\"o}ller(2015)}]{BaranMoeller2015}
{Baran, S.} and {M{\"o}ller, A.} (2015).
\newblock Joint probabilistic forecasting of wind speed and temperature using
  {B}ayesian model averaging.
\newblock \textit{Environmetrics}, {26}, 120--132.

\bibitem[{Baran et~al.(2021)Baran, Szokol and Szab{\'o}}]{baran2021truncated}
{Baran, S.}, {Szokol, P.} and {Szab{\'o}, M.} (2021).
\newblock Truncated generalized extreme value distribution-based ensemble model
  output statistics model for calibration of wind speed ensemble forecasts.
\newblock \textit{Environmetrics}, {32}, e2678.

\bibitem[{Bauer et~al.(2015)Bauer, Thorpe and Brunet}]{BauerEtAl2015}
{Bauer, P.}, {Thorpe, A.} and {Brunet, G.} (2015).
\newblock The quiet revolution of numerical weather prediction.
\newblock \textit{Nature}, {525}, 47--55.

\bibitem[{Ben~Bouall{\`e}gue et~al.(2016)Ben~Bouall{\`e}gue, Heppelmann, Theis
  and Pinson}]{BouallegueEtAl2016}
{Ben~Bouall{\`e}gue, Z.}, {Heppelmann, T.}, {Theis, S.~E.} and {Pinson, P.}
  (2016).
\newblock Generation of scenarios from calibrated ensemble forecasts with a
  dual-ensemble copula-coupling approach.
\newblock \textit{Monthly Weather Review}, {144}, 4737--4750.

\bibitem[{Benjamini and Hochberg(1995)}]{Benjamini1995}
{Benjamini, Y.} and {Hochberg, Y.} (1995).
\newblock Controlling the false discovery rate: A practical and powerful
  approach to multiple testing.
\newblock \textit{Journal of the Royal Statistical Society: Series B
  (Methodological)}, {57}, 289--300.

\bibitem[{Bergstra et~al.(2013)Bergstra, Yamins and Cox}]{bergstra2013making}
{Bergstra, J.}, {Yamins, D.} and {Cox, D.} (2013).
\newblock Making a science of model search: Hyperparameter optimization in
  hundreds of dimensions for vision architectures.
\newblock In \textit{Proceedings of the 30th International Conference on
  Machine Learning}, vol.~28. PMLR, Atlanta, Georgia, USA, 115--123.

\bibitem[{Bougeault et~al.(2010)Bougeault, Toth et~al.}]{TIGGE}
{Bougeault, P.}, {Toth, Z.} {et~al.} (2010).
\newblock The {THORPEX} interactive grand global ensemble.
\newblock \textit{Bulletin of the American Meteorological Society}, {91},
  1059--1072.

\bibitem[{Bremnes(2020)}]{bremnes2020ensemble}
{Bremnes, J.~B.} (2020).
\newblock Ensemble postprocessing using quantile function regression based on
  neural networks and {B}ernstein polynomials.
\newblock \textit{Monthly Weather Review}, {148}, 403--414.

\bibitem[{Chaloulos and Lygeros(2007)}]{ChaloulosLygeros2007}
{Chaloulos, G.} and {Lygeros, J.} (2007).
\newblock Effect of wind correlation on aircraft conflict probability.
\newblock \textit{Journal of Guidance, Control, and Dynamics}, {30},
  1742--1752.

\bibitem[{Chapman et~al.(2022)Chapman, Monache, Alessandrini, Subramanian,
  Ralph, Xie, Lerch and Hayatbini}]{Chapman2022}
{Chapman, W.~E.}, {Monache, L.~D.}, {Alessandrini, S.}, {Subramanian, A.~C.},
  {Ralph, F.~M.}, {Xie, S.-P.}, {Lerch, S.} and {Hayatbini, N.} (2022).
\newblock Probabilistic predictions from deterministic atmospheric river
  forecasts with deep learning.
\newblock \textit{Monthly Weather Review}, {150}, 215--234.

\bibitem[{Chen et~al.(2016)Chen, Duan, Houthooft, Schulman, Sutskever and
  Abbeel}]{chen2016infogan}
{Chen, X.}, {Duan, Y.}, {Houthooft, R.}, {Schulman, J.}, {Sutskever, I.} and
  {Abbeel, P.} (2016).
\newblock Info{GAN}: Interpretable representation learning by information
  maximizing generative adversarial nets.
\newblock In \textit{NIPS'16: Proceedings of the 30th International Conference
  on Neural Information Processing Systems}. Curran Associates Inc., Red Hook,
  NY, USA, 2180–2188.

\bibitem[{Clark et~al.(2004)Clark, Gangopadhyay, Hay, Rajagopalan and
  Wilby}]{clark2004schaake}
{Clark, M.}, {Gangopadhyay, S.}, {Hay, L.}, {Rajagopalan, B.} and {Wilby, R.}
  (2004).
\newblock The {S}chaake shuffle: A method for reconstructing space–time
  variability in forecasted precipitation and temperature fields.
\newblock \textit{Journal of Hydrometeorology}, {5}, 243--262.

\bibitem[{Dai and Hemri(2021)}]{DaiHemri2021}
{Dai, Y.} and {Hemri, S.} (2021).
\newblock Spatially coherent postprocessing of cloud cover ensemble forecasts.
\newblock \textit{Monthly Weather Review}, {149}, 3923--3937.

\bibitem[{Diebold and Mariano(1995)}]{Diebold1995}
{Diebold, F.~X.} and {Mariano, R.~S.} (1995).
\newblock Comparing predictive accuracy.
\newblock \textit{Journal of Business \& Economic Statistics}, {13}, 253--263.

\bibitem[{Dziugaite et~al.(2015)Dziugaite, Roy and
  Ghahramani}]{dziugaite2015training}
{Dziugaite, G.~K.}, {Roy, D.~M.} and {Ghahramani, Z.} (2015).
\newblock Training generative neural networks via {M}aximum {M}ean
  {D}iscrepancy optimization.
\newblock Preprint, \urlprefix\url{https://arxiv.org/abs/1505.03906}.

\bibitem[{Feldmann et~al.(2015)Feldmann, Scheuerer and
  Thorarinsdottir}]{FeldmannEtAl2015}
{Feldmann, K.}, {Scheuerer, M.} and {Thorarinsdottir, T.~L.} (2015).
\newblock Spatial postprocessing of ensemble forecasts for temperature using
  nonhomogeneous {G}aussian regression.
\newblock \textit{Monthly Weather Review}, {143}, 955--971.

\bibitem[{Gneiting et~al.(2007)Gneiting, Balabdaoui and
  Raftery}]{gneiting2007probabilistic}
{Gneiting, T.}, {Balabdaoui, F.} and {Raftery, A.~E.} (2007).
\newblock Probabilistic forecasts, calibration and sharpness.
\newblock \textit{Journal of the Royal Statistical Society: Series B
  (Statistical Methodology)}, {69}, 243--268.

\bibitem[{Gneiting and Raftery(2007)}]{gneiting2007strictly}
{Gneiting, T.} and {Raftery, A.~E.} (2007).
\newblock Strictly proper scoring rules, prediction, and estimation.
\newblock \textit{Journal of the American Statistical Association}, {102},
  359--378.

\bibitem[{Gneiting et~al.(2005)Gneiting, Raftery, Westveld and
  Goldman}]{gneiting2005calibrated}
{Gneiting, T.}, {Raftery, A.~E.}, {Westveld, A.~H.} and {Goldman, T.} (2005).
\newblock Calibrated probabilistic forecasting using ensemble model output
  statistics and minimum {CRPS} estimation.
\newblock \textit{Monthly Weather Review}, {133}, 1098--1118.

\bibitem[{Goodfellow et~al.(2014)Goodfellow, Pouget-Abadie, Mirza, Xu,
  Warde-Farley, Ozair, Courville and Bengio}]{goodfellow2014generative}
{Goodfellow, I.~J.}, {Pouget-Abadie, J.}, {Mirza, M.}, {Xu, B.}, {Warde-Farley,
  D.}, {Ozair, S.}, {Courville, A.} and {Bengio, Y.} (2014).
\newblock Generative adversarial nets.
\newblock In \textit{NIPS'14: Proceedings of the 27th International Conference
  on Neural Information Processing Systems - Volume 2}. MIT Press, Cambridge,
  MA, USA, 2672–2680.

\bibitem[{Gretton et~al.(2012)Gretton, Borgwardt, Rasch, Sch{\"o}lkopf and
  Smola}]{gretton2012kernel}
{Gretton, A.}, {Borgwardt, K.~M.}, {Rasch, M.~J.}, {Sch{\"o}lkopf, B.} and
  {Smola, A.} (2012).
\newblock A kernel two-sample test.
\newblock \textit{Journal of Machine Learning Research}, {13}, 723--773.

\bibitem[{Gui et~al.(2021)Gui, Sun, Wen, Tao and Ye}]{Gui2022}
{Gui, J.}, {Sun, Z.}, {Wen, Y.}, {Tao, D.} and {Ye, J.} (2021).
\newblock A review on generative adversarial networks: Algorithms, theory, and
  applications.
\newblock \textit{IEEE Transactions on Knowledge and Data Engineering} 1--1.

\bibitem[{Harris et~al.(2022)Harris, McRae, Chantry, Dueben and
  Palmer}]{HarrisEtAl2022}
{Harris, L.}, {McRae, A. T.~T.}, {Chantry, M.}, {Dueben, P.~D.} and {Palmer,
  T.~N.} (2022).
\newblock A generative deep learning approach to stochastic downscaling of
  precipitation forecasts.
\newblock Preprint, \urlprefix\url{https://arxiv.org/abs/2204.02028}.

\bibitem[{Haupt et~al.(2021)Haupt, Chapman, Adams, Kirkwood, Hosking, Robinson,
  Lerch and Subramanian}]{Haupt2021}
{Haupt, S.~E.}, {Chapman, W.}, {Adams, S.~V.}, {Kirkwood, C.}, {Hosking,
  J.~S.}, {Robinson, N.~H.}, {Lerch, S.} and {Subramanian, A.~C.} (2021).
\newblock Towards implementing artificial intelligence post-processing in
  weather and climate: proposed actions from the {O}xford 2019 workshop.
\newblock \textit{Philosophical Transactions of the Royal Society A:
  Mathematical, Physical and Engineering Sciences}, {379}, 20200091.

\bibitem[{Hess et~al.(2022)Hess, Dr{\"u}ke, Petri, Strnad and
  Boers}]{HessEtAl2022}
{Hess, P.}, {Dr{\"u}ke, M.}, {Petri, S.}, {Strnad, F.} and {Boers, N.} (2022).
\newblock Physically constrained generative adversarial networks for improving
  earth system model precipitation output.
\newblock Preprint (Version 1), available at Research Square,
  \urlprefix\url{https://doi.org/10.21203/rs.3.rs-1369622/v1}.

\bibitem[{Hu et~al.(2016)Hu, Schmeits, van Andel, Verkade, Xu, Solomatine and
  Liang}]{hu2016stratified}
{Hu, Y.}, {Schmeits, M.~J.}, {van Andel, S.~J.}, {Verkade, J.~S.}, {Xu, M.},
  {Solomatine, D.~P.} and {Liang, Z.} (2016).
\newblock A stratified sampling approach for improved sampling from a
  calibrated ensemble forecast distribution.
\newblock \textit{Journal of Hydrometeorology}, {17}, 2405--2417.

\bibitem[{Janke et~al.(2021)Janke, Ghanmi and Steinke}]{JankeEtAl2021}
{Janke, T.}, {Ghanmi, M.} and {Steinke, F.} (2021).
\newblock Implicit generative copulas.
\newblock In \textit{Advances in Neural Information Processing Systems},
  vol.~34. Curran Associates, Inc., 26028--26039.

\bibitem[{Janke and Steinke(2020)}]{janke2020probabilistic}
{Janke, T.} and {Steinke, F.} (2020).
\newblock Probabilistic multivariate electricity price forecasting using
  implicit generative ensemble post-processing.
\newblock In \textit{2020 International Conference on Probabilistic Methods
  Applied to Power Systems (PMAPS)}. IEEE, 1--6.

\bibitem[{Jordan et~al.(2019)Jordan, Krüger and Lerch}]{JordanEtAl2019}
{Jordan, A.}, {Krüger, F.} and {Lerch, S.} (2019).
\newblock Evaluating probabilistic forecasts with scoring{R}ules.
\newblock \textit{Journal of Statistical Software}, {90}, 1–37.

\bibitem[{Kingma and Ba(2014)}]{kingma2014adam}
{Kingma, D.~P.} and {Ba, J.} (2014).
\newblock {A}dam: A method for stochastic optimization.
\newblock Preprint, \urlprefix\url{https://arxiv.org/abs/1412.6980}.

\bibitem[{Lakatos et~al.(2022)Lakatos, Lerch, Hemri and
  Baran}]{LakatosEtAl2022}
{Lakatos, M.}, {Lerch, S.}, {Hemri, S.} and {Baran, S.} (2022).
\newblock Comparison of multivariate post-processing methods using global
  {ECMWF} ensemble forecasts.
\newblock Preprint, \urlprefix\url{https://arxiv.org/abs/2206.10237}.

\bibitem[{Lakshminarayanan et~al.(2017)Lakshminarayanan, Pritzel and
  Blundell}]{Lakshminarayanan2017}
{Lakshminarayanan, B.}, {Pritzel, A.} and {Blundell, C.} (2017).
\newblock Simple and scalable predictive uncertainty estimation using deep
  ensembles.
\newblock In \textit{NIPS'17: Proceedings of the 31st International Conference
  on Neural Information Processing Systems}. Curran Associates Inc., Red Hook,
  NY, USA, 6405–6416.

\bibitem[{Lang et~al.(2020)Lang, Lerch, Mayr, Simon, Stauffer and
  Zeileis}]{Lang2020}
{Lang, M.~N.}, {Lerch, S.}, {Mayr, G.~J.}, {Simon, T.}, {Stauffer, R.} and
  {Zeileis, A.} (2020).
\newblock Remember the past: a comparison of time-adaptive training schemes for
  non-homogeneous regression.
\newblock \textit{Nonlinear Processes in Geophysics}, {27}, 23--34.

\bibitem[{Lang et~al.(2019)Lang, Mayr, Stauffer and Zeileis}]{LangEtAl2019}
{Lang, M.~N.}, {Mayr, G.~J.}, {Stauffer, R.} and {Zeileis, A.} (2019).
\newblock Bivariate {G}aussian models for wind vectors in a distributional
  regression framework.
\newblock \textit{Advances in Statistical Climatology, Meteorology and
  Oceanography}, {5}, 115--132.

\bibitem[{Leinonen et~al.(2021)Leinonen, Nerini and
  Berne}]{leinonen2020stochastic}
{Leinonen, J.}, {Nerini, D.} and {Berne, A.} (2021).
\newblock Stochastic super-resolution for downscaling time-evolving atmospheric
  fields with a generative adversarial network.
\newblock \textit{IEEE Transactions on Geoscience and Remote Sensing}, {59},
  7211--7223.

\bibitem[{Lerch and Baran(2017)}]{LerchBaran2017}
{Lerch, S.} and {Baran, S.} (2017).
\newblock Similarity-based semilocal estimation of post-processing models.
\newblock \textit{Journal of the Royal Statistical Society. Series C (Applied
  Statistics)}, {66}, 29--51.

\bibitem[{Lerch et~al.(2020)Lerch, Baran, M{\"o}ller, Gro{\ss}, Schefzik, Hemri
  and Graeter}]{lerch2020simulation}
{Lerch, S.}, {Baran, S.}, {M{\"o}ller, A.}, {Gro{\ss}, J.}, {Schefzik, R.},
  {Hemri, S.} and {Graeter, M.} (2020).
\newblock Simulation-based comparison of multivariate ensemble post-processing
  methods.
\newblock \textit{Nonlinear Processes in Geophysics}, {27}, 349--371.

\bibitem[{Lerch and Polsterer(2022)}]{LerchPolsterer2022}
{Lerch, S.} and {Polsterer, K.~L.} (2022).
\newblock Convolutional autoencoders for spatially-informed ensemble
  post-processing.
\newblock International Conference on Learning Representations (ICLR) 2022 - AI
  for Earth and Space Science Workshop.,
  \urlprefix\url{https://arxiv.org/abs/2204.05102}.

\bibitem[{Lerch and Thorarinsdottir(2013)}]{LerchThorarinsdottir2013}
{Lerch, S.} and {Thorarinsdottir, T.~L.} (2013).
\newblock Comparison of non-homogeneous regression models for probabilistic
  wind speed forecasting.
\newblock \textit{Tellus A: Dynamic Meteorology and Oceanography}, {65}, 21206.

\bibitem[{Lerch et~al.(2017)Lerch, Thorarinsdottir, Ravazzolo and
  Gneiting}]{LerchEtAl2017}
{Lerch, S.}, {Thorarinsdottir, T.~L.}, {Ravazzolo, F.} and {Gneiting, T.}
  (2017).
\newblock Forecaster’s dilemma: Extreme events and forecast evaluation.
\newblock \textit{Statistical Science}, {32}, 106 -- 127.

\bibitem[{Li et~al.(2015)Li, Swersky and Zemel}]{li2015generative}
{Li, Y.}, {Swersky, K.} and {Zemel, R.} (2015).
\newblock Generative moment matching networks.
\newblock In \textit{Proceedings of the 32nd International Conference on
  Machine Learning}, vol.~37. PMLR, Lille, France, 1718--1727.

\bibitem[{Matheson and Winkler(1976)}]{matheson1976scoring}
{Matheson, J.~E.} and {Winkler, R.~L.} (1976).
\newblock Scoring rules for continuous probability distributions.
\newblock \textit{Management Science}, {22}, 1087--1096.

\bibitem[{McGovern et~al.(2019)McGovern, Lagerquist, Gagne, Jergensen, Elmore,
  Homeyer and Smith}]{mcgovern2019making}
{McGovern, A.}, {Lagerquist, R.}, {Gagne, D.~J.}, {Jergensen, G.~E.}, {Elmore,
  K.~L.}, {Homeyer, C.~R.} and {Smith, T.} (2019).
\newblock Making the black box more transparent: Understanding the physical
  implications of machine learning.
\newblock \textit{Bulletin of the American Meteorological Society}, {100},
  2175--2199.

\bibitem[{Mohamed and Lakshminarayanan(2016)}]{mohamed2016learning}
{Mohamed, S.} and {Lakshminarayanan, B.} (2016).
\newblock Learning in implicit generative models.
\newblock Preprint, \urlprefix\url{https://arxiv.org/abs/1610.03483}.

\bibitem[{M{\"o}ller et~al.(2013)M{\"o}ller, Lenkoski and
  Thorarinsdottir}]{moller2013multivariate}
{M{\"o}ller, A.}, {Lenkoski, A.} and {Thorarinsdottir, T.~L.} (2013).
\newblock Multivariate probabilistic forecasting using ensemble {B}ayesian
  model averaging and copulas.
\newblock \textit{Quarterly Journal of the Royal Meteorological Society},
  {139}, 982--991.

\bibitem[{Muschinski et~al.(2022)Muschinski, Lang, Mayr, Messner, Zeileis and
  Simon}]{MuschinskiEtAl2022}
{Muschinski, T.}, {Lang, M.~N.}, {Mayr, G.~J.}, {Messner, J.~W.}, {Zeileis, A.}
  and {Simon, T.} (2022).
\newblock Predicting power ramps from joint distributions of future wind
  speeds.
\newblock \textit{Wind Energy Science Discussions}.
\newblock Preprint, in review.

\bibitem[{Nelsen(2006)}]{Nelsen2006}
{Nelsen, R.~B.} (2006).
\newblock \textit{An Introduction to Copulas}.
\newblock 2nd ed. Springer New York, NY.

\bibitem[{Pacchiardi et~al.(2021)Pacchiardi, Adewoyin, Dueben and
  Dutta}]{PacchiardiEtAl2021}
{Pacchiardi, L.}, {Adewoyin, R.}, {Dueben, P.} and {Dutta, R.} (2021).
\newblock Probabilistic forecasting with generative networks via scoring rule
  minimization.
\newblock Preprint, \urlprefix\url{https://arxiv.org/abs/2112.08217}.

\bibitem[{Pantillon et~al.(2018)Pantillon, Lerch, Knippertz and
  Corsmeier}]{PantillonEtAl2018}
{Pantillon, F.}, {Lerch, S.}, {Knippertz, P.} and {Corsmeier, U.} (2018).
\newblock Forecasting wind gusts in winter storms using a calibrated
  convection-permitting ensemble.
\newblock \textit{Quarterly Journal of the Royal Meteorological Society},
  {144}, 1864--1881.

\bibitem[{Pedregosa et~al.(2011)Pedregosa, Varoquaux, Gramfort, Michel,
  Thirion, Grisel, Blondel, Prettenhofer, Weiss, Dubourg, Vanderplas, Passos,
  Cournapeau, Brucher, Perrot and Duchesnay}]{scikit-learn}
{Pedregosa, F.}, {Varoquaux, G.}, {Gramfort, A.}, {Michel, V.}, {Thirion, B.},
  {Grisel, O.}, {Blondel, M.}, {Prettenhofer, P.}, {Weiss, R.}, {Dubourg, V.},
  {Vanderplas, J.}, {Passos, A.}, {Cournapeau, D.}, {Brucher, M.}, {Perrot, M.}
  and {Duchesnay, E.} (2011).
\newblock Scikit-learn: Machine learning in {P}ython.
\newblock \textit{Journal of Machine Learning Research}, {12}, 2825--2830.

\bibitem[{Pennington et~al.(2014)Pennington, Socher and
  Manning}]{pennington2014glove}
{Pennington, J.}, {Socher, R.} and {Manning, C.} (2014).
\newblock {G}lo{V}e: Global vectors for word representation.
\newblock In \textit{Proceedings of the 2014 Conference on Empirical Methods in
  Natural Language Processing ({EMNLP})}. Association for Computational
  Linguistics, Doha, Qatar, 1532--1543.

\bibitem[{Perrone et~al.(2020)Perrone, Schicker and Lang}]{PerroneEtAl2020}
{Perrone, E.}, {Schicker, I.} and {Lang, M.~N.} (2020).
\newblock A case study of empirical copula methods for the statistical
  correction of forecasts of the {ALADIN}-{LAEF} system.
\newblock \textit{Meteorologische Zeitschrift}, {29}, 277--288.

\bibitem[{Petropoulos et~al.(2022)}]{PetropoulosEtAl2022}
{Petropoulos, F.} {et~al.} (2022).
\newblock Forecasting: theory and practice.
\newblock \textit{International Journal of Forecasting}, {38}, 705--871.

\bibitem[{Pinson and Girard(2012)}]{PinsonGirard2012}
{Pinson, P.} and {Girard, R.} (2012).
\newblock Evaluating the quality of scenarios of short-term wind power
  generation.
\newblock \textit{Applied Energy}, {96}, 12--20.

\bibitem[{Pinson and Messner(2018)}]{PinsonMessner2018}
{Pinson, P.} and {Messner, J.~W.} (2018).
\newblock Chapter 9 - application of postprocessing for renewable energy.
\newblock In \textit{Statistical Postprocessing of Ensemble Forecasts}.
  Elsevier, 241--266.

\bibitem[{Price and Rasp(2022)}]{PriceRasp2022}
{Price, I.} and {Rasp, S.} (2022).
\newblock Increasing the accuracy and resolution of precipitation forecasts
  using deep generative models.
\newblock In \textit{Proceedings of The 25th International Conference on
  Artificial Intelligence and Statistics}, vol. 151. PMLR, 10555--10571.

\bibitem[{Rasp and Lerch(2018)}]{rasp2018neural}
{Rasp, S.} and {Lerch, S.} (2018).
\newblock Neural networks for postprocessing ensemble weather forecasts.
\newblock \textit{Monthly Weather Review}, {146}, 3885--3900.

\bibitem[{Ravuri et~al.(2021)Ravuri, Lenc, Willson et~al.}]{ravuri2021skilful}
{Ravuri, S.}, {Lenc, K.}, {Willson, M.} {et~al.} (2021).
\newblock Skilful precipitation nowcasting using deep generative models of
  radar.
\newblock \textit{Nature}, {597}, 672--677.

\bibitem[{Schefzik(2017)}]{Schefzik2017}
{Schefzik, R.} (2017).
\newblock Ensemble calibration with preserved correlations: unifying and
  comparing ensemble copula coupling and member-by-member postprocessing.
\newblock \textit{Quarterly Journal of the Royal Meteorological Society},
  {143}, 999--1008.

\bibitem[{Schefzik and M{\"o}ller(2018)}]{schefzik2018ensemble}
{Schefzik, R.} and {M{\"o}ller, A.} (2018).
\newblock Chapter 4 - ensemble postprocessing methods incorporating dependence
  structures.
\newblock In \textit{Statistical Postprocessing of Ensemble Forecasts}.
  Elsevier, 91--125.

\bibitem[{Schefzik et~al.(2013)Schefzik, Thorarinsdottir and
  Gneiting}]{schefzik2013uncertainty}
{Schefzik, R.}, {Thorarinsdottir, T.~L.} and {Gneiting, T.} (2013).
\newblock Uncertainty quantification in complex simulation models using
  ensemble copula coupling.
\newblock \textit{Statistical Science}, {28}, 616--640.

\bibitem[{Scheuerer and Hamill(2015)}]{scheuerer2015variogram}
{Scheuerer, M.} and {Hamill, T.~M.} (2015).
\newblock Variogram-based proper scoring rules for probabilistic forecasts of
  multivariate quantities.
\newblock \textit{Monthly Weather Review}, {143}, 1321--1334.

\bibitem[{Scheuerer et~al.(2017)Scheuerer, Hamill, Whitin, He and
  Henkel}]{ScheuererEtAl2017}
{Scheuerer, M.}, {Hamill, T.~M.}, {Whitin, B.}, {He, M.} and {Henkel, A.}
  (2017).
\newblock A method for preferential selection of dates in the {S}chaake shuffle
  approach to constructing spatiotemporal forecast fields of temperature and
  precipitation.
\newblock \textit{Water Resources Research}, {53}, 3029--3046.

\bibitem[{Scheuerer and M{\"o}ller(2015)}]{ScheuererMoeller2015}
{Scheuerer, M.} and {M{\"o}ller, D.} (2015).
\newblock Probabilistic wind speed forecasting on a grid based on {E}nsemble
  {M}odel {O}utput {S}tatistics.
\newblock \textit{The Annals of Applied Statistics}, {9}, 1328--1349.

\bibitem[{Scheuerer et~al.(2020)Scheuerer, Switanek, Worsnop and
  Hamill}]{ScheuererEtAl2020}
{Scheuerer, M.}, {Switanek, M.~B.}, {Worsnop, R.~P.} and {Hamill, T.~M.}
  (2020).
\newblock Using artificial neural networks for generating probabilistic
  subseasonal precipitation forecasts over california.
\newblock \textit{Monthly Weather Review}, {148}, 3489--3506.

\bibitem[{Schuhen et~al.(2012)Schuhen, Thorarinsdottir and
  Gneiting}]{SchuhenEtAl2012}
{Schuhen, N.}, {Thorarinsdottir, T.~L.} and {Gneiting, T.} (2012).
\newblock Ensemble model output statistics for wind vectors.
\newblock \textit{Monthly Weather Review}, {140}, 3204--3219.

\bibitem[{Schulz and Lerch(2022{\natexlab{a}})}]{SchulzLerch2022ens}
{Schulz, B.} and {Lerch, S.} (2022{\natexlab{a}}).
\newblock Aggregating distribution forecasts from deep ensembles.
\newblock Preprint, \urlprefix\url{https://arxiv.org/abs/2204.02291}.

\bibitem[{Schulz and Lerch(2022{\natexlab{b}})}]{SchulzLerch2022}
{Schulz, B.} and {Lerch, S.} (2022{\natexlab{b}}).
\newblock Machine learning methods for postprocessing ensemble forecasts of
  wind gusts: A systematic comparison.
\newblock \textit{Monthly Weather Review}, {150}, 235--257.

\bibitem[{Sejdinovic et~al.(2013)Sejdinovic, Sriperumbudur, Gretton and
  Fukumizu}]{sejdinovic2013equivalence}
{Sejdinovic, D.}, {Sriperumbudur, B.}, {Gretton, A.} and {Fukumizu, K.} (2013).
\newblock Equivalence of distance-based and {RKHS}-based statistics in
  hypothesis testing.
\newblock \textit{The Annals of Statistics}, {41}, 2263--2291.

\bibitem[{Sklar(1959)}]{Sklar1959}
{Sklar, A.} (1959).
\newblock Fonctions de r\'{e}partition \`{a} $n$ dimensions et leurs marges.
\newblock \textit{Publications de l'Institut de Statistique de l'Universit\'{e}
  de Paris}, {8}, 229--231.

\bibitem[{Song and Ermon(2020)}]{song2020improved}
{Song, Y.} and {Ermon, S.} (2020).
\newblock Improved techniques for training score-based generative models.
\newblock In \textit{NIPS'20: Proceedings of the 34th International Conference
  on Neural Information Processing Systems}. 1043, Curran Associates Inc., Red
  Hook, NY, USA, 12438–12448.

\bibitem[{Sz{\'e}kely(2003)}]{szekely2003statistics}
{Sz{\'e}kely, G.~J.} (2003).
\newblock E-statistics: The energy of statistical samples.
\newblock \textit{Bowling Green State University, Department of Mathematics and
  Statistics Technical Report}, {3}, 1--18.

\bibitem[{Taillardat et~al.(2016)Taillardat, Mestre, Zamo and
  Naveau}]{TaillardatEtAl2016}
{Taillardat, M.}, {Mestre, O.}, {Zamo, M.} and {Naveau, P.} (2016).
\newblock Calibrated ensemble forecasts using quantile regression forests and
  ensemble model output statistics.
\newblock \textit{Monthly Weather Review}, {144}, 2375--2393.

\bibitem[{Thorarinsdottir and
  Gneiting(2010)}]{thorarinsdottir2010probabilistic}
{Thorarinsdottir, T.~L.} and {Gneiting, T.} (2010).
\newblock Probabilistic forecasts of wind speed: ensemble model output
  statistics by using heteroscedastic censored regression.
\newblock \textit{Journal of the Royal Statistical Society: Series A
  (Statistics in Society)}, {173}, 371--388.

\bibitem[{Van~Schaeybroeck and Vannitsem(2015)}]{vanSchaeybroeckVannitsem2015}
{Van~Schaeybroeck, B.} and {Vannitsem, S.} (2015).
\newblock Ensemble post-processing using member-by-member approaches:
  theoretical aspects.
\newblock \textit{Quarterly Journal of the Royal Meteorological Society},
  {141}, 807--818.

\bibitem[{Vannitsem et~al.(2021)Vannitsem, Bremnes, Demaeyer, Evans, Flowerdew,
  Hemri, Lerch, Roberts, Theis, Atencia, Bouallègue, Bhend, Dabernig, Cruz,
  Hieta, Mestre, Moret, Plenković, Schmeits, Taillardat, den Bergh,
  Schaeybroeck, Whan and Ylhaisi}]{vannitsem2021statistical}
{Vannitsem, S.}, {Bremnes, J.~B.}, {Demaeyer, J.}, {Evans, G.~R.}, {Flowerdew,
  J.}, {Hemri, S.}, {Lerch, S.}, {Roberts, N.}, {Theis, S.}, {Atencia, A.},
  {Bouallègue, Z.~B.}, {Bhend, J.}, {Dabernig, M.}, {Cruz, L.~D.}, {Hieta,
  L.}, {Mestre, O.}, {Moret, L.}, {Plenković, I.~O.}, {Schmeits, M.},
  {Taillardat, M.}, {den Bergh, J.~V.}, {Schaeybroeck, B.~V.}, {Whan, K.} and
  {Ylhaisi, J.} (2021).
\newblock Statistical postprocessing for weather forecasts: Review, challenges,
  and avenues in a big data world.
\newblock \textit{Bulletin of the American Meteorological Society}, {102},
  E681--E699.

\bibitem[{Wilks(2015)}]{wilks2015multivariate}
{Wilks, D.~S.} (2015).
\newblock Multivariate ensemble {M}odel {O}utput {S}tatistics using empirical
  copulas.
\newblock \textit{Quarterly Journal of the Royal Meteorological Society},
  {141}, 945--952.

\bibitem[{Worsnop et~al.(2018)Worsnop, Scheuerer, Hamill and
  Lundquist}]{WorsnopEtAl2018}
{Worsnop, R.~P.}, {Scheuerer, M.}, {Hamill, T.~M.} and {Lundquist, J.~K.}
  (2018).
\newblock Generating wind power scenarios for probabilistic ramp event
  prediction using multivariate statistical post-processing.
\newblock \textit{Wind Energy Science}, {3}, 371--393.

\bibitem[{Zhou(2022)}]{Zhou2022}
{Zhou, B.} (2022).
\newblock Interpreting generative adversarial networks for interactive image
  generation.
\newblock In \textit{xxAI - Beyond Explainable AI: International Workshop, Held
  in Conjunction with ICML 2020, July 18, 2020, Vienna, Austria, Revised and
  Extended Papers}. Springer International Publishing, Cham, 167--175.

\bibitem[{Ziel and Berk(2019)}]{ZielBerk2019}
{Ziel, F.} and {Berk, K.} (2019).
\newblock Multivariate forecasting evaluation: On sensitive and strictly proper
  scoring rules.
\newblock Preprint, \urlprefix\url{https://arxiv.org/abs/1910.07325}.

\end{thebibliography}

\newpage

\begin{appendices}
	
	\section{Strategies for generating CGM ensembles}\label{appendix:ensembling}
	
	\begin{figure}[!htb]
		\centering
		\includegraphics[width=0.65\textwidth]{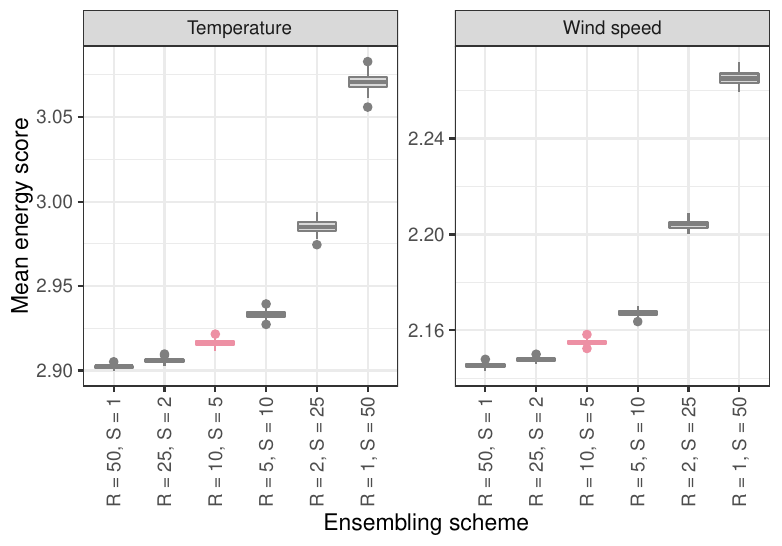}
		\caption{Boxplots of mean energy scores of different CGM ensembling schemes comprised of $R$ model runs producing $S$ samples each. The mean scores are computed across the 100 sub-sampled station datasets, and the boxplots summarize the variability across 100 repetitions of the ensembling procedure by randomly selecting $R$ models runs and $S$ generated samples. Our ensembling strategy from the main text is marked in red. We consider only the case $D=10$ here.}
		\label{fig_cgm_ensemble}
	\end{figure}
	
	While generating ensembles of CGM forecasts improves predictive performance (see Section \ref{sec:cgm-sample-size}), there exist a variety of strategies and configurations to 
	obtain the final $\nout$ samples. In the CGM implementation utilized in the main text, we repeated the model estimation 10 times and generated 5 samples each. Here, we investigate the effects of alternative strategies to obtain  $\nout = 50$ samples by estimating 50 (or 25, 5, 2, 1) CGM model(s) and generating 1 (or 2, 10, 25, 50) sample(s) each.
	To that end, we repeat the CGM estimation 50 times and generate 100 samples each. From this large set of CGM predictions, we then randomly select a set of $R \in \{1,2,5,10,25,50\}$ models runs and $S \in \{50,25,10,5,2,1\}$ corresponding samples such that $R\cdot S = \nout = 50$. For each of these combinations, we compute the mean energy score over the 100 repetitions of the experiment (i.e., the 100 sets of sub-sampled stations) described in the main paper. This procedure is repeated 100 times. 
	
	Figure \ref{fig_cgm_ensemble} shows boxplots of the corresponding mean scores and clearly indicates that increasing the number of runs $R$ substantially improves the predictive performance, with the best results obtained for $R = 50$ runs producing a single sample each. These improvements in predictive performance need to be balanced with against the increased computational costs. For example, increasing $R$ from 10 to 50 roughly leads to a 5-fold increase in computational costs, since the computational costs of generating additional samples from an estimated CGM is negligible. However, the mean energy scores typically improve by less than 1\%. The configuration we selected ($R = 10, S = 5$) represents a reasonable compromise between computational costs and forecast performance. 
	
	Note that multiple sources of randomness contribute to the variability of the performance of the CGM ensembles. Additional results (not shown) indicate that the variability is dominated by the randomness introduced via repeated model runs due to the stochastic training procedure, in comparison to which the randomness due to the sample generation process is negligible.
	
	\section{On the role of additional inputs for the CGM predictive performance}\label{appendix:inputs}
	
	\begin{figure}[h]
		\centering
		\includegraphics[width=\textwidth]{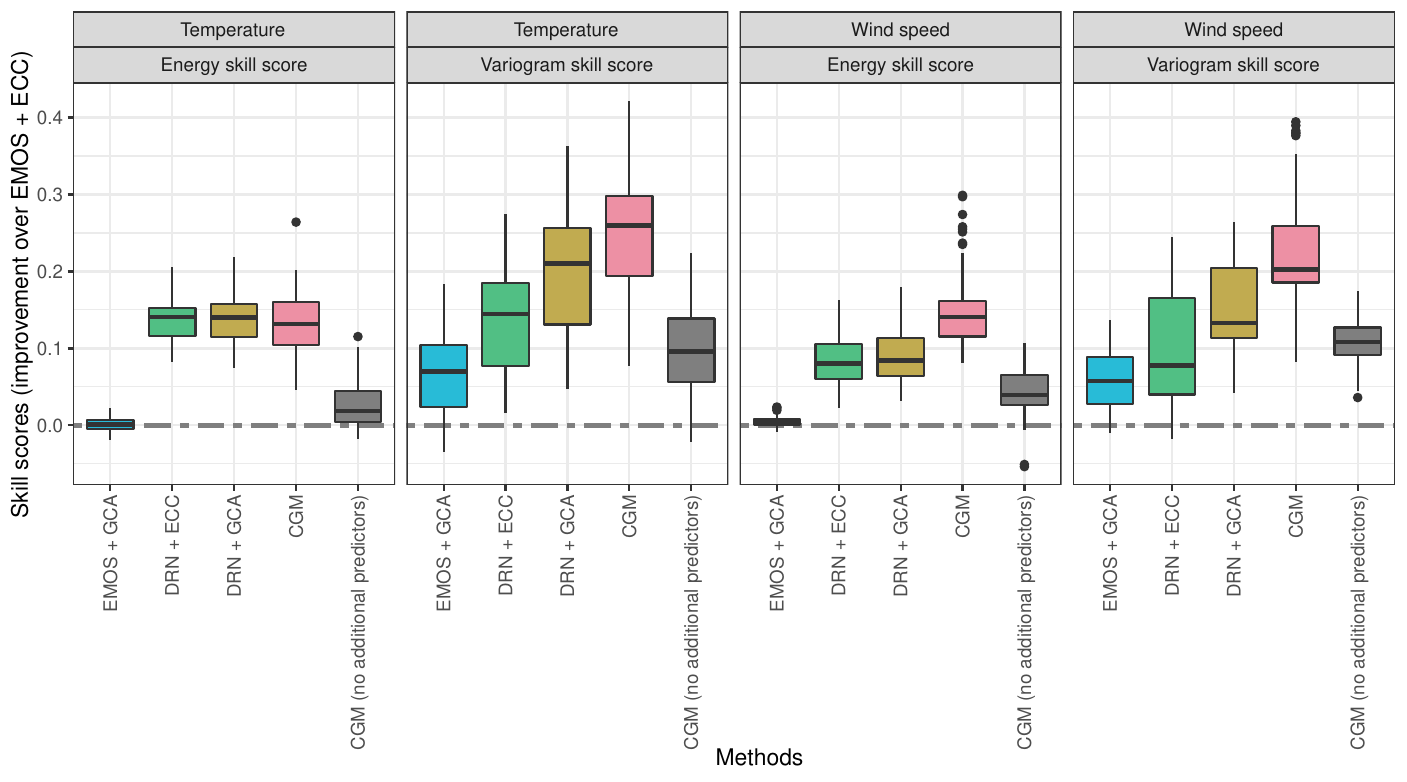}
		\caption{Boxplots of energy skill scores and variogram skill scores of different multivariate post-processing methods analogous to Figures \ref{fig_tem_mvscores} and \ref{fig_ws_mvscores} for $D=10$, but including a CGM variant without additional inputs. EMOS+ECC is used as reference forecast throughout.}
		\label{fig_cgm_noaddinput}
	\end{figure}
	
	To assess the importance of incorporating additional input features for the CGM performance, Figure \ref{fig_cgm_noaddinput} includes a CGM variant which only uses ensemble forecasts of the target variable as input. While this CGM variant has access to the same information as the EMOS+ECC and EMOS+GCA models, it generally shows superior predictive performance, indicating improvements of the CGM approach beyond utilizing additional input features only. While the CGM variant without additional inputs typically fails to achieve forecast performance comparable to the DRN-based models, it on average outperforms the DRN+ECC model for wind speed in terms of the VS.

\end{appendices}

\end{document}